\documentstyle[aaspp4,12pt,flushrt,axodraw]{article}

\begin{document} 

\def\la{\mathrel{\mathpalette\fun <}}
\def\ga{\mathrel{\mathpalette\fun >}}

\def\fun#1#2{\lower3.6pt\vbox{\baselineskip0pt\lineskip.9pt
        \ialign{$\mathsurround=0pt#1\hfill##\hfil$\crcr#2\crcr\sim\crcr}}}
                 

\title{Cosmological Perturbations: Entering the Non-Linear Regime}

\author{ Rom\'an Scoccimarro\footnote{{\it present address}: CITA, 
McLennan Physical Labs, 60 St George Street, Toronto, ON M5S 3H8.} }
\affil{Department of Physics and Enrico Fermi Institute, 
University of Chicago, Chicago, IL~60637, and NASA/Fermilab 
Astrophysics Center, Fermi National Accelerator 
Laboratory, P.O.Box 500, Batavia, IL 60510}


\begin{abstract}

We consider one-loop corrections (non-linear corrections {\it beyond} 
leading order) to the bispectrum and skewness 
of cosmological density  fluctuations induced by gravitational 
evolution, focusing on the case of 
 Gaussian initial conditions and  scale-free initial power spectra,
 $P(k) \propto k^{n}$. 
As  has been established by comparison with numerical simulations, 
tree-level (leading order) perturbation theory describes these 
quantities at the largest scales. One-loop  
perturbation theory provides a tool to probe the transition to the 
non-linear regime on smaller scales.
In this work, we find that, as a function of spectral index $n$, 
the one-loop bispectrum 
follows a pattern analogous to  that of the one-loop power 
spectrum, which shows a change in behavior at a ``critical index''
$n_{c} \approx -1.4$, where non-linear corrections vanish. 
For the bispectrum, for $n \la n_{c}$, one-loop 
corrections increase the configuration dependence of the leading 
order contribution; 
 for $n \ga n_{c}$, one-loop corrections tend to cancel the 
configuration dependence of the tree-level bispectrum, in  
agreement with known results from  $n=-1$  numerical simulations. A 
similar situation is shown to hold for the Zel'dovich approximation,
where $n_{c} \approx -1.75$. 
Using dimensional regularization, we obtain explicit analytic expressions  
for the one-loop bispectrum for $n=-2$ initial power spectra,
 for both the exact dynamics of gravitational 
instability and  the Zel'dovich approximation. We also compute the 
skewness factor, including local averaging of the density field, for $n=-2$:   
$S_3(R) = 4.02 + 3.83 \ \sigma^2_{G}(R)$ for gaussian smoothing and 
$S_3(R) = 3.86 + 3.18 \ \sigma^2_{TH}(R)$ for top-hat smoothing,
where $\sigma^2(R)$ is the variance of the density field fluctuations 
smoothed over a window of radius $R$. Comparison  
 with fully non-linear numerical simulations  implies that, for $n < 
 -1$,  
one-loop  perturbation theory can extend our  
understanding of nonlinear clustering  down  
to scales where the transition to the stable clustering regime begins. 

\end{abstract}


\section{Introduction}
\label{cha:intro}

There is growing evidence that the large-scale structure of the Universe 
grew via gravitational instability from small  
primordial fluctuations in the matter density. 
For realistic models of structure formation, the initial 
spectrum of perturbations is such that at large scales,  fluctuations 
are small and reflect the primordial spectrum. The variance of 
density fluctuations, $\sigma^{2} (R)$, is a decreasing function of scale 
$R$. At small scales,  $\sigma ^{2} (R)$ is large enough  that 
non-linear effects become important. There are therefore two limiting 
regimes characterized by the value of $\sigma ^{2} (R)$: the {\it linear 
regime} 
at large scales, where $\sigma ^{2} (R) \ll 1$, and the {\it non-linear 
regime} at 
small scales, where $\sigma ^{2} (R) \gg 1$. The boundary between these two 
regimes defines a length scale, the correlation length $R_{0}$, where $\sigma ^{2} 
(R_{0}) =1$. Because of gravitational instability, $R_{0}$ grows with 
time and therefore a given scale eventually becomes non-linear under 
time evolution. 

At early epochs, the growth of density perturbations can 
be described by linear perturbation theory, provided that the linear power 
spectrum $P(k)$  falls off less steeply 
than $k^4$ for small $k$ (Zel'dovich 1965, Peebles 1974, Peebles
\& Groth 1976). 
In the linear regime, perturbation Fourier 
modes evolve independently of one another, conserving the statistical properties  
of the primordial fluctuations. In particular, if the primordial fluctuations 
are Gaussian random fields, they remain Gaussian in linear theory. In 
this case, the statistical properties of the density and velocity fields 
are completely determined by the 
two-point correlation function or the power spectrum.

When the fluctuations become non-linear, 
coupling between different Fourier modes becomes important,  
inducing non-trivial correlations that modify
the statistical properties of the cosmological fields. For Gaussian initial
conditions, this causes the appearance of higher-order reduced correlations, 
which constitute independent statistics that can be measured 
in observational data and numerical simulations, even when the departure 
from the linear regime is small. 

Non-linear cosmological perturbation 
theory provides a theoretical
framework for the calculation of the induced higher-order 
correlation functions in the {\it weakly non-linear regime}, defined 
by scales $R$ such that $\sigma (R) \la 1$.
At large scales, leading order (tree-level) 
perturbation theory gives the first non-vanishing contribution, and 
has been used to understand the generation of higher order 
correlations in gravitational instability. Comparison with fully 
non-linear numerical simulations has shown this approach to be very 
successful (Juszkiewicz, Bouchet \& Colombi 1993, 
Bernardeau 1994b, \L okas et al. 1995, Gazta\~{n}aga \& Baugh
1995, Baugh, Gazta\~{n}aga \& Efstathiou 1995).

As one approaches smaller scales, however, next to leading order (loop) 
corrections to the tree-level results are expected to become 
important. The question then arises of whether our understanding of 
non-linear clustering can be extended from the largest scales into the 
transition region to the non-linear regime. This motivates us to 
consider one-loop cosmological 
perturbation theory. In previous work (Scoccimarro \& Frieman
1996b, hereafter SF2), we showed 
that for scale-free initial conditions, $P(k) \propto k^{n}$,
 without too much small-scale power (spectral index $n< -1$), 
one can understand the evolution of the power spectrum down to scales 
where it begins to go over to the strongly non-linear 
stable clustering  regime. Therefore, it is 
interesting to consider one-loop corrections to the higher order 
correlation functions as well, to see if one can gain similar 
understanding of non-linear clustering on intermediate scales.

In this work we concentrate on  one-loop corrections to
the three-point function of density perturbations in Fourier 
space, also known as the bispectrum, and its one-point counterpart, 
the skewness. These are interesting quantities for several reasons. 
On the theoretical side, the bispectrum is the lowest order 
correlation function which, for Gaussian initial conditions,
vanishes in the linear regime; 
its structure therefore reflects truly non-linear properties of the 
matter distribution. Furthermore, as the lowest order correlation 
function which depends on the vector character of its arguments, it gives 
direct physical information on the anisotropic structures and flows 
generated by gravitational instability.
Observationally, the configuration dependence of the 
tree-level bispectrum has been put forward as 
a promising statistic to study the important but poorly 
understood issue of bias (Fry 1994), i.e., the degree to which 
luminous objects in the universe such as galaxies are fair tracers of the 
underlying density field. It is therefore important to see how 
further non-linear effects (which are inevitably present  in 
observational studies) alter this 
configuration dependence, to check whether  
one can still disentangle nonlinear evolution from  bias.

We focus on Gaussian initial conditions and
scale-free initial power spectra, $P(k) \propto k^{n}$. 
In addition to  mathematical simplicity, primordial 
Gaussian fluctuations have a broad physical motivation 
and are predicted by the simplest inflationary models. 
Although the linear power spectrum 
for the Universe is not scale-free (on both observational and 
theoretical grounds), scale-free spectra are very useful 
approximations over limited ranges of wavenumber $k$. They also have 
the advantage of yielding analytic closed form results and giving rise 
to self-similar evolution  of the statistical properties of cosmological 
fields for spatially flat universes (Davis \& Peebles 1977,
Peebles 1980). In particular, 
self-similarity is a powerful aid towards a physical understanding of 
non-linear clustering and in many realistic models of structure 
formation we expect approximate self-similar evolution over a 
restricted range of length and time-scales (Efstathiou et al. 1988).

While the agreement between tree-level perturbation theory and 
numerical simulations in the weakly non-linear regime is well 
established, it was not until recent years that N-body simulations have 
been able to reliably follow the transition of higher order statistics
into the non-linear 
regime. In this regard, for scale-free initial power spectra, the 
bispectrum in numerical simulations 
has been shown by Fry, Melott \& Shandarin (1993,1995) 
 to depart from the 
tree-level perturbative results at scales comparable to  
the correlation length, as expected if next to leading order 
corrections are present. Similarly, for the skewness of 
the density field, and for higher-order cumulants as well, 
deviations from the leading order calculations have 
been reported in the literature (Bouchet \& Hernquist 1992, 
Lucchin et al. 1994, Juszkiewicz et al. 1995, Hivon et al. 1995,
Colombi, Bouchet \& Hernquist 1996).  
We  therefore consider it appropriate to extend the leading order 
calculations 
to one-loop, in order  
to understand better the limitations of the tree-level results and see the 
extent to which one can improve the agreement of perturbation theory 
with fully non-linear 
numerical simulations.

This paper is organized as follows. In Section~\ref{cha:dynstat} 
we discuss the cosmological fluid 
equations of motion and their solution within the 
framework of perturbation theory.  For recent reviews 
of perturbation theory see Bernardeau (1996), Bouchet (1996),
Juszkiewicz \& Bouchet (1996);  
approximation methods in gravitational clustering are reviewed 
by Sahni \& Coles (1996). Section~\ref{cha:statdia} reviews 
the diagrammatic approach to perturbation theory and the 
self-similarity properties of statistical quantities derived from it. 
The main results of this work are presented in 
Section~\ref{cha:results}, where we consider for the first time 
one-loop corrections to the bispectrum and skewness including 
smoothing effects. We compare the latter with results from numerical 
simulations; a similar comparison for the bispectrum will be 
presented elsewhere (Scoccimarro et al 1996). 
Section~\ref{cha:conc} contains our conclusions. 
Auxiliary material is consider in the Appendices.

\section{Dynamics and Perturbation Theory}
\label{cha:dynstat}

\subsection{Equations of Motion}
\label{sec:eqsmotion}

The equations of motion relevant to gravitational instability
describe conservation of mass and 
momentum and the Poisson equation 
for a self-gravitating perfect fluid with zero 
pressure in a homogeneous and isotropic universe (Peebles 1980):


\begin{equation}
	{\partial \delta({\bf x},\tau) \over{\partial \tau}} + \nabla \cdot  \{
	[1+\delta{({\bf x},\tau)}] {\bf v}({\bf x},\tau) \} =  0
	\label{continuity},
\end{equation}
\begin{equation}
     {\partial {\bf v}({\bf x},\tau) \over{\partial \tau}} + {{\cal H}(\tau)}\ 
     {\bf v}({\bf x},\tau) + [{\bf 
	v}({\bf x},\tau) \cdot \nabla] {\bf v}({\bf x},\tau) =  - \nabla 
	\Phi({\bf x},\tau)
	\label{euler},
\end{equation}	
\begin{equation}
	\nabla^2 \Phi({\bf x},\tau)  = {3\over 2} \Omega {\cal H}^2(\tau) 
	\delta{({\bf x},\tau)}
	\label{poisson}
\end{equation}

\noindent\noindent Here, ${\bf x}$ denotes comoving spatial coordinates, 
$\tau = \int dt/a$ is the  conformal time, $a(\tau)$ is the cosmic scale factor,
the density contrast $\delta{({\bf x},\tau)} \equiv
\rho{({\bf x},\tau)}/ \bar \rho  - 1$, with $\bar \rho(\tau)$ the mean density 
of matter, $ {\bf v} \equiv d{\bf x}/d\tau $  represents the velocity field 
fluctuations about the Hubble flow,  $ {\cal H}\equiv {d\ln a /{d\tau}}=Ha$
is the conformal expansion rate,  
 $\Phi$ is the gravitational potential due to the 
density fluctuations, and the density parameter $\Omega = \bar \rho/\rho_c = 
8\pi G \bar \rho a^2/3{\cal H}^2$.  We take the velocity field to be 
irrotational, so it can be completely described by its divergence $ 
\theta \equiv \nabla \cdot {\bf v }$.
We will refer to Eqs.~(\ref{continuity})-(\ref{poisson}) 
as the ``exact dynamics" (ED), to make a distinction with 
the modified dynamics introduced by non-linear approximations 
such as the Zel'dovich approximation and the Local Lagrangian 
approximation to be discussed later (see 
Appendices~\ref{app:ZA} and~\ref{app:lla}).
Equations~(\ref{continuity})-(\ref{poisson}) hold in an arbitrary 
homogeneous and isotropic background Universe which evolves according to the
Friedmann equations; henceforth, for simplicity we assume an Einstein-de Sitter 
background, $\Omega =1$, with vanishing 
cosmological constant, for which  $a \propto \tau^2$ and $3 \Omega 
{\cal H}^2 /2 = 6/\tau^2$. 

 Taking the divergence of Equation~(\ref{euler}) and Fourier transforming the 
resulting equations of motion we get:

\label{eqsfourier}
\begin{equation}
	{\partial \tilde{\delta}({\bf k},\tau) \over{\partial \tau}} + 
	\tilde{\theta}({\bf k},\tau) = - \int d^3k_1 \int d^3k_2 \delta_D({\bf k} 
	- {\bf k}_1 - {\bf k}_2) \alpha({\bf k}, 
	{\bf k}_1) \tilde{\theta}({\bf k}_1,\tau) 
	\tilde{\delta}({\bf k}_2,\tau)
	\label{ddtdelta},
\end{equation}
\begin{eqnarray}
	{\partial \tilde{\theta}({\bf k},\tau) \over{\partial \tau}} &+& 
	{\cal H}(\tau)\ \tilde{\theta}({\bf k},\tau) + {3\over 2}  {\cal 
	H}^2(\tau) \tilde{\delta}({\bf k},\tau) = \nonumber \\
      &-& \int d^3k_1 \int d^3k_2 \delta_D({\bf k} 
	- {\bf k}_1 - {\bf k}_2) \beta({\bf k}, {\bf k}_1, {\bf k}_2) 
	\tilde{\theta}({\bf k}_1,\tau) \tilde{\theta}({\bf k}_2,\tau)
	\label{ddttheta},
\end{eqnarray}

\noindent ($\delta_D$ denotes the  three-dimensional Dirac delta distribution),  
where the functions 

\begin{equation}
	\alpha({\bf k}, {\bf k}_1) \equiv {{\bf k} \cdot {\bf 
k}_1 \over{ k_1^2}}, \ \ \ \ \ \beta({\bf k}, {\bf 
k}_1, {\bf k}_2) \equiv {k^2 ({\bf k}_1 \cdot {\bf k}_2 )\over{2 k_1^2 
k_2^2}}
	\label{albe}
\end{equation}

\noindent encode the non-linearity of the evolution (mode coupling) and come from the 
non-linear terms in the continuity equation~(\ref{continuity}) and the Euler 
equation~(\ref{euler}) respectively.

\subsection{Perturbation Theory Solutions}
\label{sec:recrel}

We focus on a statistical description of 
cosmological perturbations:   
we are  interested in  correlation functions of the fields 
$\tilde{\delta}({\bf k},\tau)$ and $ 
\tilde{\theta}({\bf k},\tau)$ (i.e., the ensemble 
average of products of these fields). Ensemble averaging
 effectively introduces a 
new parameter into the problem, the variance of the density 
fluctuations $\sigma^{2} \equiv < \delta^{2} >$, 
which controls the transition from the linear ($\sigma^{2}~\ll~1$) to the 
non-linear regime ($\sigma^{2}~\gg~1$).
We consider 
perturbations about the linear solution, effectively treating the 
variance of the linear fluctuations as a small parameter.
In this case, Eqs.~(\ref{ddtdelta})-(\ref{ddttheta})  can be formally  solved 
via a perturbative expansion, 

\begin{equation}
	\tilde{\delta}({\bf k},\tau) = \sum_{n=1}^{\infty} a^n(\tau) 
	\delta_n({\bf k}),\ \ \ \ \ \tilde{\theta}({\bf k},\tau) = 
	{\cal H}(\tau) \sum_{n=1}^{\infty} a^n(\tau) \theta_n({\bf k})
	\label{ptansatz},
\end{equation}

\noindent where only the fastest growing mode at each order 
is taken into account. At small $a$,  
the series are dominated by their first terms, and since  $\theta_1({\bf k}) = 
-\delta_1({\bf k}) $ from the continuity equation, $\delta_1({\bf k})$ 
completely characterizes the linear fluctuations. The equations of 
motion~(\ref{ddtdelta})-(\ref{ddttheta}) determine $\delta_n({\bf k}) $ and 
$\theta_n({\bf k})$  in terms of the linear fluctuations,

\label{solu}
\begin{equation}
	\delta_n({\bf k}) = \int d^3q_1 \ldots \int d^3q_n \delta_D({\bf k} - 
	{\bf q}_1 - \ldots - {\bf q}_n) F_n^{(s)}({\bf q}_1,  \ldots  ,{\bf q}_n) 
	\delta_1({\bf q}_1) \ldots \delta_1({\bf q}_n)
	\label{ec:deltan},
\end{equation}
\begin{equation}
		\theta_n({\bf k}) = - \int d^3q_1 \ldots \int d^3q_n \delta_D({\bf k} - 
	{\bf q}_1 - \ldots - {\bf q}_n) G_n^{(s)}({\bf q}_1,  \ldots  ,{\bf q}_n) 
	\delta_1({\bf q}_1) \ldots \delta_1({\bf q}_n)
	\label{ec:thetan},
\end{equation}


\noindent where $F_n^{(s)}$ and $G_n^{(s)}$ are symmetric homogeneous 
functions  with degree zero of the 
wave vectors \{${\bf q}_1,  \ldots  ,{\bf q}_n $\}. They 
are constructed from the fundamental mode coupling functions $\alpha({\bf k}, 
{\bf k}_1)$ and $\beta({\bf k}, {\bf k}_1, {\bf k}_2)$ according to 
the recursion relations ($n \geq 2$, see Goroff et al. (1986) or
Jain \& Bertschinger (1994) for a 
derivation):

\label{ec:recrel}
\begin{eqnarray}
 F_n({\bf q}_1,  \ldots  ,{\bf q}_n) &=& \sum_{m=1}^{n-1}
 { G_m({\bf q}_1,  \ldots  ,{\bf q}_m) \over{(2n+3)(n-1)}} 
 \Bigl[(2n+1) \alpha({\bf k},{\bf k}_1)  F_{n-m}({\bf q}_{m+1},  \ldots  
 ,{\bf q}_n) \nonumber \\ 
&+& 2 \beta({\bf k},{\bf k}_1, {\bf k}_2)  G_{n-m}({\bf q}_{m+1},  
 \ldots  ,{\bf q}_n) \Bigr]
 \label{ec:Fn},
\end{eqnarray}
\begin{eqnarray}
	 G_n({\bf q}_1,  \ldots  ,{\bf q}_n) &=& \sum_{m=1}^{n-1}
 { G_m({\bf q}_1,  \ldots  ,{\bf q}_m) \over{(2n+3)(n-1)}} 
 \Bigl[3 \alpha({\bf k},{\bf k}_1)  F_{n-m}({\bf q}_{m+1},  \ldots  
 ,{\bf q}_n) \nonumber \\
&+& 2n \beta({\bf k},{\bf k}_1, {\bf k}_2)  G_{n-m}({\bf q}_{m+1},  
 \ldots   ,{\bf q}_n) \Bigr]
	\label{ec:Gn},
\end{eqnarray}

\noindent (where ${\bf k}_1 \equiv {\bf q}_1 + \ldots 
+ {\bf q}_m$,  ${\bf k}_2 \equiv 
{\bf q}_{m+1} + \ldots + {\bf q}_n$,  ${\bf k} \equiv {\bf k}_1 +{\bf 
k}_2$, and $F_1= G_1 \equiv 1$) and the symmetrization 
procedure:

\label{ec:symm}
\begin{equation}
 F_n^{(s)}({\bf q}_1,  \ldots  ,{\bf q}_n) = {1\over{n!}}\sum_{\pi }
 F_n({\bf q}_{\pi (1)},  \ldots  ,{\bf q}_{\pi (n)})
 \label{ec:Fns},
\end{equation}
\begin{equation}
	 G_n^{(s)}({\bf q}_1,  \ldots  ,{\bf q}_n) ={1\over{n!}} \sum_{\pi }
  G_n({\bf q}_{\pi (1)},  \ldots  ,{\bf q}_{\pi (n)})
	\label{ec:Gns},
\end{equation}

\noindent where the sum is taken over all the permutations $\pi $ of the set 
$\{1, \ldots ,n\}$.

\section{Statistics and Diagrammatics}
\label{cha:statdia}

\subsection{Diagrammatic Expansion of Statistical Quantities }
\label{sec:diagrams}

The starting point for a statistical description of fluctuations in 
cosmology is the ``Fair Sample Hypothesis'' (Peebles 1980, 
Bertschinger 1992). 
This asserts 
that fluctuations can be described by statistically homogeneous and isotropic 
random fields (so that our Universe is a random realization from a 
statistical ensemble) and that within the accessible part of the Universe 
there are many independent samples that can be considered to approximate 
a statistical ensemble, so that spatial averages are equivalent to ensemble 
averages (``ergodicity'').
In this work we focus on the non-linear evolution of the 
three-point cumulant 
of the density field, the bispectrum $ B({\bf k}_1,{\bf k}_2,\tau)$, 
and its 1-point counterpart, the 
skewness factor $S_3(R,\tau) $. These 
are defined respectively by:

\begin{equation}
	\Big< \tilde{\delta}({\bf k}_1,\tau) \tilde{\delta}
({\bf k}_2,\tau) \tilde{\delta}({\bf k}_3,\tau) \Big>_c = 
\delta_D({\bf k}_1+{\bf k}_2+{\bf k}_3) \ B({\bf k}_1,{\bf k}_2,\tau)
	\label{bispectrum},
\end{equation} 

\noindent and 

\begin{equation}
	S_3(R,\tau) = \frac{1}{\sigma^4(R,\tau)} 
	\int B({\bf k}_1,{\bf k}_2,\tau) \ W(k_1 R)
	W(k_2 R) W(|{\bf k}_1 +{\bf k}_2| R) \ d^3 k_{1}  d^3 k_{2}
	\label{skewness},
\end{equation}

\noindent where the angle brackets denote ensemble averaging , the 
subscript ``c'' stands for the connected contribution 
(see below), and $\sigma^2(R,\tau)$ is the variance of the density field 
fluctuations:

\begin{equation}
	\sigma^2(R,\tau) = \int P(k,\tau) \ W^2(k R) \ d^3 k 
	= \Big< \delta^2(R,\tau) \Big> 
	\label{variance}.
\end{equation}

\noindent Here the power spectrum $P(k,\tau)$ is defined by

\begin{equation}
	\Big< \tilde{\delta}({\bf k},\tau) \tilde{\delta}
({\bf k}',\tau)\Big>_c = \delta_D({\bf k}+{\bf k}') P(k,\tau)
	\label{powerspectrum},
\end{equation} 

\noindent and therefore

\begin{equation}
	S_3(R,\tau) = \frac{\Big< \delta^3(R,\tau) \Big>_c 
}{\Big< \delta^2(R,\tau) \Big>^{2}}
	\label{skewness2}.
\end{equation}

\noindent Here $W(k R)$ is the Fourier transform of the 
window function,  which we take to be either 
a top-hat (TH) or a Gaussian (G), 

\begin{equation}
	W_{\rm TH} (u)  =   \frac{3}{u^3}\Big[ \sin (u) - u \cos (u) \Big]
	\label{WTH} ,
\end{equation}

\begin{equation}
	W_{\rm G} (u)  =  \exp (-u^2 /2)
	\label{WG}.
\end{equation}

\noindent It is convenient to define the  
hierarchical amplitude $Q$ as follows (Fry \& Seldner 1982, Fry 1984):

\begin{equation}
    Q \equiv \frac{ B({\bf k}_1,{\bf k}_2,\tau)}{ P(k_1,\tau) P(k_2,\tau) +
     P(k_2,\tau) P(k_3,\tau) +  P(k_3,\tau) P(k_1,\tau) }
	\label{q},
\end{equation} 

\noindent which has the desirable property that it is scale and time 
independent to 
lowest order (tree-level) 
in non-linear perturbation theory. In a pure hierarchical model, $Q$ 
would be a fixed constant, independent of configuration and of 
the power spectrum $P(k,\tau)$ as well.

We are interested in calculating the non-linear evolution of 
these statistical quantities from Gaussian initial conditions
in the weakly non-linear regime, $\sigma(R) \la 1$.  
A systematic framework for  calculating  correlations of cosmological 
fields in perturbation theory has been formulated using
diagrammatic techniques (Goroff et al. 1986, Wise 1988,
Scoccimarro \& Frieman 1996 (SF1) , SF2).  In this 
approach, contributions to $p$-point cumulants 
of the density field come from 
connected diagrams with $p$ external (solid) lines and $r=p-1, p, \dots$ 
internal (dashed) lines.  The perturbation expansion
leads to a collection of diagrams at each order, the leading order being 
tree-diagrams, the next to leading order  1-loop diagrams and so on.  
In each diagram, external lines represent the spectral components 
of the fields we are interested in (e.g., $\delta({\bf k},\tau)$). 
Each internal line is labeled by a wave-vector 
that is integrated over, and represents a linear power spectrum  
$P_{11}(q,\tau)$. Vertices of order $n$ (i.e., where $n$ internal lines 
join) represent an $n^{\rm th}$ order perturbative solution $\delta_n$, 
and momentum conservation is imposed at each 
vertex.

\begin{figure}

\begin{center}\begin{picture}(330,100)(0,0)

\small

\ArrowLine(0,50)(20,50)
\Vertex(20,50){3}
\DashArrowLine(20,50)(70,50){3}
\Vertex(70,50){3}
\ArrowLine(70,50)(90,50)
\Text(45,57)[]{${\bf k}$}
\Text(45,0)[]{$(P_{11})$}

\Text(100,50)[]{+}
\Text(110,50)[]{$\Bigg[$}

\ArrowLine(120,50)(140,50)
\Vertex(140,50){3}
\DashCurve{(140,50)(165,60)(190,50)}{3}
\DashCurve{(140,50)(165,40)(190,50)}{3}
\DashArrowLine(164,60)(166,60){3}
\DashArrowLine(164,40)(166,40){3}
\Vertex(190,50){3}
\ArrowLine(190,50)(210,50)
\Text(165,67)[]{${\bf k}-{\bf q}$}
\Text(165,33)[]{${\bf q}$}
\Text(165,0)[]{$(P_{22})$}

\Text(220,50)[]{+}

\ArrowLine(230,50)(250,50)
\Vertex(250,50){3}
\DashArrowArcn(250,60)(10,0,180){3}
\DashArrowArcn(250,60)(10,180,360){3}
\DashArrowLine(250,50)(300,50){3}
\Vertex(300,50){3}
\ArrowLine(300,50)(320,50)
\Text(250,77)[]{${\bf q}$}
\Text(275,57)[]{${\bf k}$}
\Text(275,0)[]{$(P_{13})$}
\Text(330,50)[]{$\Bigg]$}

\end{picture}\end{center}
\normalsize 
\caption{ Diagrams  for the power spectrum
 up to one-loop. See Eqs.~(\protect \ref{P22}) and (\protect \ref{P13})
for diagram amplitudes.}
\label{fig2}

\end{figure}
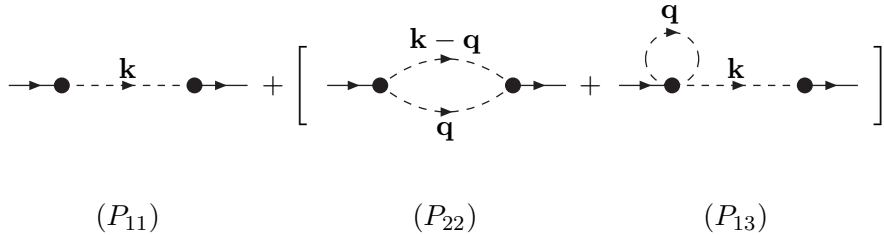

We can write the loop 
expansion for the power spectrum up to one-loop corrections as 

\begin{equation}
	P(k,\tau) = P^{(0)}(k,\tau) + P^{(1)}(k,\tau) + \ldots 
	\label{Ploopexp},
\end{equation}

\noindent where the superscript $(n)$ denotes an $n$-loop contribution, 
the tree-level ($0$-loop) contribution is just the linear spectrum,

\begin{equation}
	P^{(0)}(k,\tau) =  P_{11}(k,\tau)  
	\label{P^(0)},
\end{equation}

\noindent with $a^2(\tau) \langle \delta_1({\bf k}) \delta_1({\bf k}')\rangle_c 
= \delta_D({\bf k}+{\bf k}') P_{11}(k,\tau)$, 
and the 1-loop contribution consists of two terms,

\begin{equation}
	P^{(1)}(k,\tau) = P_{22}(k,\tau) + P_{13}(k,\tau) 
	\label{P^(1)},
\end{equation}

\noindent with (see Fig.~\ref{fig2}):

\begin{eqnarray}
	 P_{22}(k,\tau) & \equiv  &  2 \int [F_2^{(s)}({\bf k}-{\bf q},{\bf q}) ]^2
	 P_{11}(|{\bf k}-{\bf q}|,\tau)  P_{11}(q,\tau)  d^3q
	\label{P22},  \\
	 P_{13}(k,\tau)  & \equiv &   6  \int 
	 F_3^{(s)}({\bf k},{\bf q},-{\bf q}) P_{11}(k,\tau)	
	 P_{11}(q,\tau)  d^3q
	\label{P13}.
\end{eqnarray}

\noindent Here $P_{ij}$ denotes the 
amplitude given by the above rules
for a connected diagram representing the contribution from  
$\langle \delta_i  \delta_j \rangle_c$ to the power spectrum. We have 
assumed Gaussian initial conditions, for which  
$P_{ij}$ vanishes if $i+j$ is odd.

For the smoothed variance  we write

\label{sigxiexp}
\begin{equation}
	\sigma^2(R)  = 
	\sigma_\ell^2(R) \Big( 1 + s^{(1)} \ \sigma_\ell^2(R)
	+ \ldots \Big)
	\label{siexp},
\end{equation} 

\noindent where $\sigma_\ell^2(R)$  denotes the 
variance  in linear theory 
(given by (\ref{variance})  with $P=P_{11}$); 
the dimensionless 1-loop amplitude is 

\begin{equation}
	 s^{(1)}(R)  \equiv  \frac{1}{\sigma_\ell^4(R)}
	 \int P^{(1)}(k,\tau) \ W^2(k R) \ d^3 k
	\label{s1}. 
\end{equation} 

To characterize the degree of non-linear evolution when including 
one-loop corrections to the power spectrum and bispectrum, it is convenient to  
define a physical scale from the linear power spectrum, 
the correlation length $R_0$,  as
the scale where the smoothed linear variance is unity,

\begin{equation}
         \sigma^2_\ell(R_0)  = \int d^{3}k \ P_{11}(k,\tau) \ W^{2}(k R_{0})
         \equiv  1
        \label{R0}.
\end{equation}

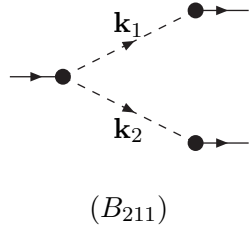
\begin{figure}
\small

\begin{center}\begin{picture}(100,100)(0,0)

\ArrowLine(0,50)(20,50)
\Vertex(20,50){3}
\DashArrowLine(20,50)(70,75){3}
\DashArrowLine(20,50)(70,25){3}
\Vertex(70,75){3}
\Vertex(70,25){3}
\ArrowLine(70,75)(90,75)
\ArrowLine(70,25)(90,25)
\Text(45,70)[]{${\bf k}_1$}
\Text(45,30)[]{${\bf k}_2$}
\Text(45,0)[]{$( B_{211})$}

\end{picture}\end{center}
\normalsize 

\caption{Tree-level diagram  for the bispectrum. This diagram plus 
its 2 permutations over external momenta generates the tree-level 
bispectrum.  
The corresponding 
amplitudes are given by Eq.~(\protect \ref{Btree}).}
\label{fig3}
\end{figure}

The loop expansion for the bispectrum reads:

\begin{equation}
B({\bf k}_1,{\bf k}_2,\tau) = B^{(0)}({\bf k}_1,{\bf k}_2,\tau) + 
B^{(1)}({\bf k}_1,{\bf k}_2,\tau) + \ldots 
	\label{Bloopexp},
\end{equation}

\noindent where the tree-level part is given by a single diagram in 
second order perturbation theory (see Fig.~\ref{fig3})
 plus its permutations over external 
momenta (recall that ${\bf k}_1 + {\bf k}_2
+{\bf k}_3 \equiv 0$):

\begin{eqnarray}
B^{(0)}({\bf k}_1,{\bf k}_2,\tau) &\equiv& 
2  P_{11}(k_1,\tau)  P_{11}(k_2,\tau) F_2^{(s)}({\bf k}_1,{\bf k}_2) +
2  P_{11}(k_2,\tau)  P_{11}(k_3,\tau)
\nonumber \\	 & & \times  F_2^{(s)}({\bf k}_2,{\bf k}_3)  + 
2  P_{11}(k_3,\tau)  P_{11}(k_1,\tau) F_2^{(s)}({\bf k}_3,{\bf k}_1)
	\label{Btree}.
\end{eqnarray}

\noindent The one-loop contribution consists of four distinct 
diagrams 
involving up to fourth-order solutions:

\begin{equation}
B^{(1)}({\bf k}_1,{\bf k}_2,\tau) \equiv B_{222}({\bf k}_1,{\bf 
k}_2,\tau) + B_{321}^{I}({\bf k}_1,{\bf k}_2,\tau) +
 B_{321}^{II}({\bf k}_1,{\bf k}_2,\tau) + B_{411}({\bf k}_1,{\bf k}_2,\tau)
	\label{B1loop},
\end{equation}

\noindent where:

\label{1loopB}
\begin{eqnarray}
	 B_{222} & \equiv & 8  \int d^{3}q P_{11}(q,\tau) 
	 F_2^{(s)}(-{\bf q},{\bf q}+{\bf k}_1)  P_{11}(|{\bf q}+{\bf k}_1|,\tau)
	 F_2^{(s)}(-{\bf q}- {\bf k}_1,{\bf q}-{\bf k}_2) \nonumber \\
	 & & \ \ \  \times P_{11}(|{\bf q}-{\bf k}_2|,\tau)
	 F_2^{(s)}({\bf k}_2-{\bf q},{\bf q}) 
	\label{B222}, \\
	B_{321}^I & \equiv & 6 P_{11}(k_3,\tau)  \int d^{3}q P_{11}(q,\tau)
	 F_3^{(s)}(-{\bf q},{\bf q}-{\bf k}_2,-{\bf k}_3 )  
	 P_{11}(|{\bf q}-{\bf k}_2|,\tau) \nonumber \\
	 & &   \ \times 
	F_2^{(s)}({\bf q},{\bf k}_2-{\bf q}) + {\rm permutations}
	\label{B321I}, \\
	B_{321}^{II} & \equiv & 6 
	 P_{11}(k_2,\tau)  P_{11}(k_3,\tau)  F_2^{(s)}({\bf k}_2,{\bf k}_3) 
	\int d^{3}q P_{11}(q,\tau)
    F_3^{(s)}({\bf k}_3,{\bf q},-{\bf q}) \nonumber \\
	 & &   + {\rm permutations}
	\label{B321II}, \\
	B_{411} & \equiv & 12  P_{11}(k_2,\tau)  P_{11}(k_3,\tau)
	\int d^{3}q P_{11}(q,\tau)
    F_4^{(s)}({\bf q},-{\bf q},-{\bf k}_2,-{\bf k}_3) \nonumber \\
	 & &   + {\rm permutations}
	\label{B411}.
\end{eqnarray} 

\begin{figure}

\begin{center}\begin{picture}(420,100)(0,0)

\small

\ArrowLine(0,50)(20,50)
\Vertex(20,50){3}
\DashArrowLine(70,75)(20,50){3}
\DashArrowLine(20,50)(70,25){3}
\DashArrowLine(70,25)(70,75){3}
\Vertex(70,75){3}
\Vertex(70,25){3}
\ArrowLine(90,75)(70,75)
\ArrowLine(90,25)(70,25)
\Text(45,70)[]{${\bf q}$}
\Text(38,30)[]{${\bf q}+{\bf k}_{1}$}
\Text(53,50)[]{${\bf q}-{\bf k}_{2}$}
\Text(10,57)[]{${\bf k}_{1}$}
\Text(80,82)[]{${\bf k}_{2}$}
\Text(80,32)[]{${\bf k}_{3}$}
\Text(45,0)[]{$( B_{222})$}

\Text(100,50)[]{+}

\ArrowLine(110,50)(130,50)
\Vertex(130,50){3}

\DashCurve{(130,50)(151,72)(180,75)}{3}
\DashArrowLine(151,72)(150,71){3}
\DashCurve{(130,50)(159,54)(180,75)}{3}
\DashArrowLine(159,54)(158,53){3}

\DashArrowLine(180,25)(130,50){3}
\Vertex(180,75){3}
\Vertex(180,25){3}
\ArrowLine(200,75)(180,75)
\ArrowLine(200,25)(180,25)
\Text(144,75)[]{${\bf q}$}
\Text(155,30)[]{${\bf k}_3$}
\Text(177,50)[]{${\bf k}_2-{\bf q}$}
\Text(120,57)[]{${\bf k}_{1}$}
\Text(190,82)[]{${\bf k}_{2}$}
\Text(190,32)[]{${\bf k}_{3}$}
\Text(155,0)[]{$(B_{321}^I)$}

\Text(210,50)[]{+}

\ArrowLine(220,50)(240,50)
\Vertex(240,50){3}
\DashArrowLine(290,75)(240,50){3}
\DashArrowLine(287,25)(240,50){3}
\DashArrowArcn(290,35)(10,0,180){3}
\DashArrowArcn(290,35)(10,180,360){3}
\Vertex(290,75){3}
\Vertex(290,25){3}
\ArrowLine(310,75)(290,75)
\ArrowLine(310,25)(290,25)
\Text(265,70)[]{${\bf k}_2$}
\Text(265,30)[]{${\bf k}_3$}
\Text(290,52)[]{${\bf q}$}
\Text(230,57)[]{${\bf k}_{1}$}
\Text(300,82)[]{${\bf k}_{2}$}
\Text(310,32)[]{${\bf k}_{3}$}
\Text(265,0)[]{$(B_{321}^{II})$}

\Text(320,50)[]{+}

\ArrowLine(330,50)(350,50)
\Vertex(350,50){3}
\DashArrowLine(400,75)(353,50){3}
\DashArrowLine(400,25)(353,50){3}
\DashArrowArcn(350,60)(10,0,180){3}
\DashArrowArcn(350,60)(10,180,360){3}
\Vertex(400,75){3}
\Vertex(400,25){3}
\ArrowLine(420,75)(400,75)
\ArrowLine(420,25)(400,25)
\Text(375,70)[]{${\bf k}_2$}
\Text(375,30)[]{${\bf k}_3$}
\Text(350,77)[]{${\bf q}$}
\Text(336,57)[]{${\bf k}_{1}$}
\Text(410,82)[]{${\bf k}_{2}$}
\Text(410,32)[]{${\bf k}_{3}$}
\Text(375,0)[]{$(B_{411})$}

\end{picture}\end{center}
\normalsize 

\caption{One-loop diagrams  for the bispectrum. 
The corresponding 
amplitudes are given in Eqs.~(\protect \ref{B222}) through
(\protect \ref{B411}).}
\label{fig4}
\end{figure}
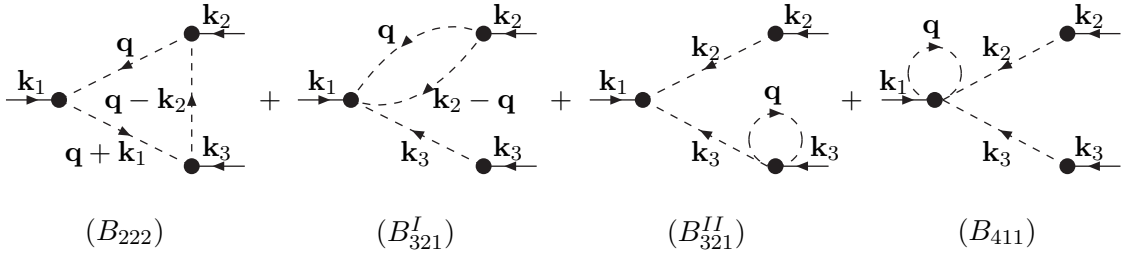


\noindent  For the hierarchical amplitude $Q$ (see Eq.~(\ref{q})), 
the loop expansion  yields:
 
\begin{equation}
    Q \equiv \frac{ B^{(0)}({\bf k}_1,{\bf k}_2,\tau) +
	B^{(1)}({\bf k}_1,{\bf k}_2,\tau) + \ldots}
    { \Sigma^{(0)}({\bf k}_1,{\bf k}_2,\tau) + 
	\Sigma^{(1)}({\bf k}_1,{\bf k}_2,\tau) + \ldots }
	\label{Qexpand},
\end{equation} 

\noindent where:

\begin{equation}
    \Sigma^{(0)}({\bf k}_1,{\bf k}_2,\tau) 
	\equiv  P_{11}(k_1,\tau) P_{11}(k_2,\tau) +
     P_{11}(k_2,\tau) P_{11}(k_3,\tau) +  P_{11}(k_3,\tau) P_{11}(k_1,\tau) 
	\label{sigma0},
\end{equation} 

\noindent and:

\begin{equation}
    \Sigma^{(1)}({\bf k}_1,{\bf k}_2,\tau) 
	\equiv  P^{(0)}(k_1,\tau) P^{(1)}(k_2,\tau) +
	 {\rm permutations} 
	\label{sigma1}.
\end{equation} 

\noindent For large scales, it is possible to expand 
$Q \equiv  Q^{(0)} + Q^{(1)} +~\ldots $, which gives:

\begin{equation}
    Q^{(0)} \equiv \frac{ B^{(0)}({\bf k}_1,{\bf k}_2,\tau)}
    { \Sigma^{(0)}({\bf k}_1,{\bf k}_2,\tau) }
	\label{qtree},
\end{equation} 
 
\begin{equation}
    Q^{(1)}   \equiv   \frac{ B^{(1)}({\bf k}_1,{\bf k}_2,\tau)
    - Q^{(0)} \Sigma^{(1)}({\bf k}_1,{\bf k}_2,\tau)}
    { \Sigma^{(0)}({\bf k}_1,{\bf k}_2,\tau) } 
      \equiv  \tilde{Q}^{(1)} - Q^{(0)} \ 
     \frac{ \Sigma^{(1)}({\bf k}_1,{\bf k}_2,\tau) }
    {\Sigma^{(0)}({\bf k}_1,{\bf k}_2,\tau)}
     \label{q1l}.
\end{equation}

\noindent Note that $Q^{(1)}$ depends on the normalization 
of the linear power spectrum, and 
its amplitude increases  with time evolution.
On the other hand,  from Equations~(\ref{Btree}) and (\ref{qtree}) 
it follows that $Q^{(0)}$ is 
independent of time and normalization (Fry 1984). 
Furthermore, for scale-free initial conditions, $P_{11}(k) \propto 
k^{n}$, $Q^{(0)}$ is also independent of overall scale.
For the particular case of
equilateral configurations ($k_{1}=k_{2}=k_{3}$ and $\hat{k}_{i} \cdot 
\hat{k}_{j}=-0.5$ for all pairs), $Q^{(0)}$ is independent of spectral index as 
well, $Q^{(0)}_{EQ}=4/7$. In general, for scale-free initial power spectra, 
$Q^{(0)}$ depends on configuration shape through, e.g., the ratio 
$k_{1}/k_{2}$ and the angle  $\theta$  defined by $\hat{k}_{1} \cdot 
\hat{k}_{2} =\cos{\theta}$. Note that we also defined $\tilde{Q}^{(1)}$ in
Eq.~(\ref{q1l}) which denotes the one-loop correction to the bispectrum
normalized by the tree-level quantity $\Sigma^{(0)}$; this  
will be useful in order to assess the behavior of the one-loop bispectrum
with spectral index.

\begin{figure}[t!]
\centering
\centerline{\epsfxsize=11. truecm \epsfysize=10. truecm 
\epsfbox{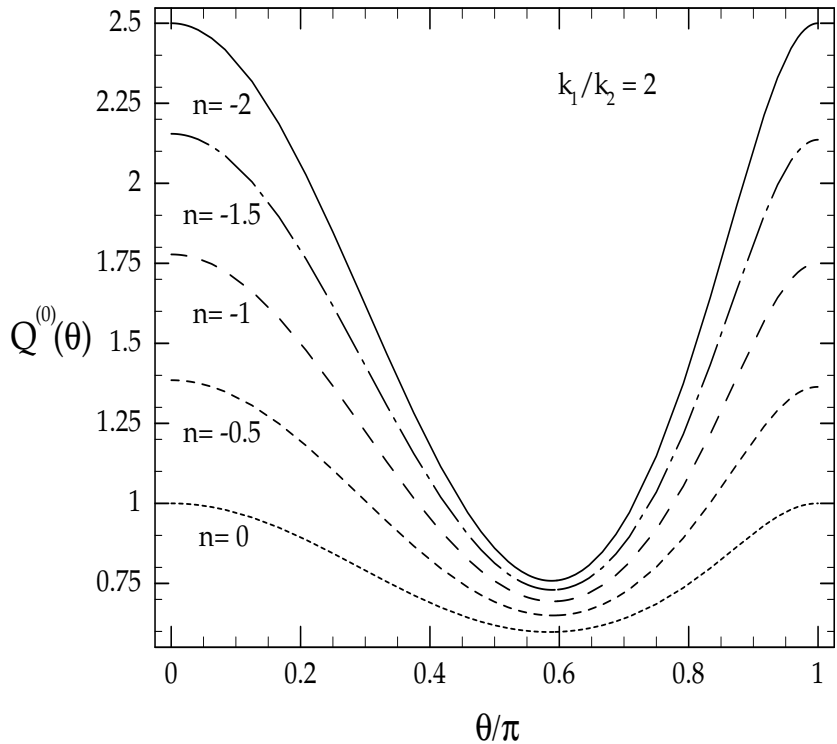}}

\caption{The tree-level hierarchical amplitude $Q^{(0)}$
 for triangle configurations given by $k_{1}/k_{2}=2$ 
as  a function of the angle $\theta$ ($\hat{k}_{1} \cdot 
\hat{k}_{2} =\cos{\theta}$). The different curves correspond to 
spectral indices $n=-2,-1.5,-1,-0.5,0$ (from top to 
bottom). See also Fry (1994)}
\label{figbitree}
\end{figure}

Figure~\ref{figbitree} shows $Q^{(0)}$ for the triangle configuration 
given by  $k_{1}/k_{2}=2$ as a function of $\theta$ for different 
spectral indices. The configuration dependence of $Q^{(0)}$ is 
remarkably insensitive to other cosmological parameters, such as the 
density parameter $\Omega$ and the cosmological 
constant (Fry 1994) (see also Hivon et al. (1995)). In fact, since bias 
between the galaxies and the underlying density field is known to 
change this configuration dependence (Fry \& Gazta\~{n}aga
1993), 
measurements of 
the hierarchical amplitude $Q$ in galaxy surveys 
could provide a measure of bias which is insensitive to  other poorly 
known cosmological parameters (Fry 1994), unlike the usual 
determination from peculiar velocities which has a degeneracy with  
the density parameter $\Omega$. 

The configuration dependence of $Q^{(0)}$ comes from the second
order perturbation theory kernel $F_2^{(s)}$ (see
Eqs.~(\ref{qtree}) and (\ref{Btree})) and 
can be  understood in physical terms 
as follows. From the recursion relations given in
Eq.~(\ref{ec:Fn}), we can write:

\begin{equation}
 F_2^{(s)}({\bf k}_1,{\bf k}_2)= \frac{5}{14} \ 
    \Big[  \alpha({\bf k}, {\bf k}_1) + \alpha({\bf k},{\bf
    k}_2) \Big] + \frac{2}{7} \ \beta({\bf k},{\bf k}_1,{\bf k}_2)
     \label{F2phys},
\end{equation} 

\noindent where $ {\bf k} \equiv {\bf k}_1 + {\bf k}_2$ (see
Eq.~(\ref{albe}) for definitions of the mode-coupling functions
$\alpha$ and $\beta$). The
terms in square brackets contribute a constant
term, independent of configuration,
 coming from the $\theta \times \delta$ term in
the equations of motion, plus  terms which depend on
configuration and describe gradients of the density field
in the direction of the flow (i.e., the term ${\bf v} \cdot
\nabla \delta$ in the continuity equation). Similarly, the last
term in Eq.~(\ref{F2phys}) contributes configuration dependent
terms which come from gradients of the velocity divergence in
the direction of the flow (due to the term $({\bf v} \cdot \nabla)
{\bf v}$ in Euler's equation). Therefore, the configuration
dependence of the bispectrum reflects the anisotropy of
structures and flows generated by gravitational instability. 
The enhancement of correlations 
for  collinear wavevectors ($\theta =0, \pi$) in
Figure~\ref{figbitree}, reflects the fact that gravitational
instability generates density and velocity divergence gradients
which are mostly parallel to the flow. Upon ensemble averaging,
which by ergodicity 
corresponds to weighting configurations by their number frequency,
this leads to a predominance of correlations in nearly collinear
configurations.  
The dependence on the
spectrum is also easy to understand: models with more
large-scale power (smaller spectral indices $n$) 
give rise to anisotropic structures and flows with larger
coherence length, which upon ensemble averaging leads to a more
anisotropic bispectrum.
We will see in the next Section 
that this physical picture provides some 
insight into the behavior of one-loop corrections.

 The loop expansion for the  skewness factor  gives (SF1):

\begin{equation}
S_{3}(R) \equiv \frac{S_{3}^{(0)} + \tilde{S}_{3}^{(1)} \sigma^2(R) + 
\ldots }{1 + 2 s^{(1)} \sigma^2(R) + \ldots} 
	\label{S3expand},
\end{equation}

\noindent where:

\begin{equation}
S_{3}^{(0)}(R) \equiv \frac{1}{\sigma_{\ell}^{4}(R)} 
\int d^{3}k_{1} d^{3}k_{2} B^{(0)}({\bf k}_1,{\bf k}_2) 
\ W(k_1 R) W(k_2 R) W(|{\bf k}_1 +{\bf k}_2| R)
	\label{S3tree},
\end{equation}

\begin{equation}
\tilde{S}_{3}^{(1)}(R)  \equiv  \frac{1}{\sigma_{\ell}^{6}(R)} 
\int d^{3}k_{1} d^{3}k_{2} B^{(1)}({\bf k}_1,{\bf k}_2) 
\ W(k_1 R) W(k_2 R) W(|{\bf k}_1 +{\bf k}_2| R) 
	\label{S3loop}.
\end{equation}

\noindent For large scales, the expansion in Eq.~(\ref{S3expand}) can be 
rewritten as $S_{3} \equiv S_{3}^{(0)} + 
 S_{3}^{(1)} \sigma^{2} + \ldots$, where:

\begin{equation}
S_{3}^{(1)}(R)  \equiv  \tilde{S}_{3}^{(1)}(R)  
	  - 2 \ s^{(1)}(R) \ S_{3}^{(0)}(R)
	\label{S3one}.
\end{equation}

\noindent The tree-level skewness has been thoroughly 
studied (Goroff et al. 1986, Juszkiewicz et al. 1993, 
Bernardeau 1992, Bernardeau 1994, \L okas et al. 1995) 
including smoothing effects for both top-hat and Gaussian smoothing. 
The result for scale-free initial power spectra is:

\begin{equation}
S_{3}^{(0)} = \frac{34}{7} - (n+3)
	\label{S3treeTH},
\end{equation}

\noindent for top-hat smoothing (Bernardeau 1992b, Juszkiewicz
et al. 1993), and:

\begin{equation}
S_{3}^{(0)} = 3  \ {_{2}{\rm F}_{1}} \Big(\frac{n+3}{2},\frac{n+3}{2},
\frac{3}{2},\frac{1}{4}\Big) - \Big(n+\frac{8}{7}\Big) \  
 {_{2}{\rm F}_{1}} \Big(\frac{n+3}{2},\frac{n+3}{2},
\frac{5}{2},\frac{1}{4}\Big) 
	\label{S3treeG},
\end{equation}

\noindent for Gaussian smoothing (\L okas et al. 1995,
Matsubara 1994),
 where 
${_{2}{\rm F}_{1}}$ denotes a hypergeometric function. More specifically, 
$S_{3}^{(0)}= 4.02, \ 3.71, \ 3.47, \ 3.28, \ 3.14$ for $n=-2,-1.5, -1, -0.5, 
0$ for Gaussian smoothing. See Appendix~\ref{app:ZA} for the 
corresponding 
results in the Zel'dovich approximation.

\subsection{Self-Similarity and Perturbation Theory}
\label{sec:selfsimilarity}

Since there is 
no preferred scale in the dynamics of a self-gravitating  
pressureless perfect fluid in an 
Einstein-de Sitter universe,  
Eqs.~(\ref{continuity})-(\ref{poisson}) admit self-similar 
solutions (Peebles 1980). 
This means that correlation functions of the cosmological fields 
should scale with a self-similarity variable, given appropriate initial 
conditions: knowing a statistical quantity   
at a given time completely specifies its evolution. For Gaussian 
initial conditions and scale-free power spectra, 
one can define a physical scale $R_{0}$, 
the correlation length (see Eq.~(\ref{R0})), 
which obeys $R_{0} \propto a^{2/(n+3)}$ in linear theory (and in 
general if the non-linear power spectrum evolves self-similarly). 
Statistical quantities in linear perturbation theory 
evolve self-similarly with $R_{0}$, e.g.,

\begin{equation}
	 R_0^{-3} P^{(0)}(k, \tau) \equiv  {\cal P}^{(0)} ( k R_0)
	\label{pss2},
\end{equation}
 
\begin{equation}
	 R_0^{-6} B^{(0)}({\bf k}_{1},{\bf k}_{2},\tau) \equiv  
	 {\cal B}^{(0)}({\bf k}_{1}R_0,{\bf k}_{2}R_0)
	\label{biss2}.
\end{equation}

\noindent When loop corrections are taken into account, however, 
self-similarity may be broken by the appearance of new 
scales  required by infrared 
and ultraviolet divergences in the loop integrations. 
In fact, one may consider a linear 
 power spectrum $P_{11}(k,\tau)$ given by a
truncated power-law,

\begin{equation}
 P_{11}(k,\tau) \equiv \left\{ \begin{array}{ll} 
                               A \ a^2(\tau) \ k^n & \mbox{if $\epsilon \leq 
                               k \leq k_c$}, \\
                               0 & \mbox{otherwise},
                               \end{array}
                       \right. 
\label{P11}
\end{equation}

\noindent where $A$ is a normalization constant; the infrared and 
ultraviolet cutoffs $\epsilon$ and 
$k_c$ are imposed in order to regularize the  loop
integrations (SF1). In a cosmological N-body simulation, they would 
correspond roughly to the inverse comoving box size and lattice spacing (or 
interparticle separation) respectively.

In the absence of the cutoffs, the spectrum 
(\ref{P11}) would be scale-free. 
The introduction of fixed (time-independent) 
cutoff scales $\epsilon$ and $k_c$ 
in the linear power spectrum (\ref{P11}) breaks 
self-similarity, because they do not scale with the 
self-similarity variable $k R_{0}$. The 
extent to which one can take the limits $\epsilon 
\rightarrow 0$ and $k_c \rightarrow \infty$ will determine whether 
one recovers self-similar scaling for
the statistical properties of the density field.  
Infrared divergences, regulated by $\epsilon$, 
arise in individual diagrams when $n \leq -1$, due to the divergence of 
the rms velocity field at large scales (Jain \& Bertschinger
1996, SF1). 
These divergences are just a 
kinematical effect and cancel when the sum over diagrams is done, as a  
consequence of the Galilean invariance of the equations of 
motion (SF1).

Ultraviolet divergences, on the other hand, arise due to small-scale 
power, and become more severe as $n$ increases. In fact, for $n \geq 
-1$, one-loop corrections to the power spectrum break 
self-similarity (SF2) 

\begin{equation}
	{\cal P}(k R_0) = \frac{(k R_0)^{n}}{2 \pi 
	\ \Gamma \left( \frac{n+3}{2} \right)}  - \frac{61 \ (k R_0)^{2n+3}}
	{315 \pi  (n+1) \  \Gamma^2 \left( \frac{n+3}{2} \right) } \ \left( 
	\frac{k}{k_c} \right)^{\eta}
	\label{ssp1lngeq0},
\end{equation}
   
\noindent where $\eta = -(n+1)$ is an exponent which measures the deviation 
from self-similar scaling, and the self-similarity breaking factor 
becomes a logarithm when $n=-1$. 

We have done a similar calculation for the ``tadpole'' diagram 
($B_{321}^{II}$ and $B_{411}$) contributions to the bispectrum 
and found that the same self-similarity breaking factors appear in 
this case. For the other contributions, the explicit calculation is 
not possible to do  analytically, but it can be checked by 
numerical integration that the full one-loop bispectrum breaks 
self-similarity for $n \geq -1$. 
For equilateral configurations, this calculation can be 
summarized by  the one-loop result for $n >-1$:

\begin{equation}
	{\cal B}^{(1)}(k R_0,k R_{0}) = b_{n}  \ (k R_0)^{3n+3}  \left( 
	\frac{k}{k_c} \right)^{\eta}
	\label{ssb1lngeq0},
\end{equation}

\noindent with $b_{n}$ a finite $n$-dependent constant factor  and 
logarithmic terms breaking self-similarity as $n 
\rightarrow -1$. This breaking of self-similar evolution, not seen 
in numerical simulations, is  a feature of the perturbative 
approach; it comes from large loop momenta due to increasing small 
scale power as $n$ increases. 
We therefore do not expect our results 
to be physical as $n \rightarrow -1$ from below. On the other hand, 
for $-3 < n<-1$, one-loop corrections to the bispectrum 
scale as $(k R_0)^{3n+3}$ and preserve self-similar evolution. In this 
case, as was already considered for the power spectrum in SF2, 
it is more convenient to regularize the individual 
one-loop bispectrum diagrams by using 
dimensional regularization (see Appendix~\ref{app:dr}), 
which effectively takes the 
limits $\epsilon \rightarrow 0$ and $k_c \rightarrow \infty$.  We now turn 
to the results of this calculation.

\section{One-Loop Results: Entering the Non-Linear Regime}
\label{cha:results}

\subsection{Bispectrum}
\label{sec:bi}

We now  consider one-loop corrections to the bispectrum for initial 
power-law spectra $P_{11}(k) \propto k^{n}$ 
with spectral index $-3 < n<-1$. In this case, the 
resulting bispectrum obeys self-similarity and, based on previous 
results for the power spectrum (SF2),  the perturbative approach 
is expected to give a good description of the transition to the 
nonlinear regime. 
Due to statistical homogeneity and isotropy, the bispectrum $B({\bf 
k}_{1},{\bf k}_{2},\tau)$ in the scaling regime ($ \epsilon \ll k_i \ll 
k_c$) only depends on time,  
the quantities $k_{1}$, $k_{2}$, 
and the angle $\theta$ ( $\hat{k}_{1}\cdot \hat{k}_{2} \equiv 
\cos{\theta}$). In order to display the analytic results, however,
it is more convenient to trade the 
variable $\theta$ for the third side of the triangle, $k_{3} = |{\bf k}_1 + 
{\bf k}_2 |$. 
Let $B^{(1)}({\bf k}_{1},{\bf k}_{2}) \equiv A^{3} a^{6}
\pi^{3} \ b^{(1)}(k_1,k_2,k_3)$, with
${\bf k}_1 + {\bf k}_2
+{\bf k}_3 \equiv 0$. Then, using the results of Appendix~\ref{app:dr},
and summing over diagrams according to the results in 
Section~\ref{sec:diagrams}, 
the one-loop correction to the bispectrum for $n=-2$ reads:

\begin{eqnarray}
	 b^{(1)}(k_1,k_2,k_3) & = & -{{30279}\over {34496\,{{k_1}^3}}} - 
{{2635\,{{k_1}^2}}\over {51744\,{{k_2}^5}}} - 
{{37313\,k_1}\over {206976\,{{k_2}^4}}} + 
{{38431}\over {68992\,k_1\,{{k_2}^2}}}    
	\nonumber   \\
	 &  & + 
{{233\,{{k_1}^6}}\over {8624\,{{k_2}^4}\,{{k_3}^5}}} - 
{{16517\,{{k_1}^5}}\over 
    {362208\,{{k_2}^3}\,{{k_3}^5}}} + 
{{197\,{{k_1}^4}}\over {7392\,{{k_2}^2}\,{{k_3}^5}}} - 
{{78691\,{{k_1}^3}}\over {275968\,k_2\,{{k_3}^5}}}
	\nonumber  \\
	 &  & -     
{{23\,{{k_1}^5}}\over 
    {103488\,{{k_2}^4}\,{{k_3}^4}}} + 
{{9791\,{{k_1}^4}}\over 
    {206976\,{{k_2}^3}\,{{k_3}^4}}} + 
{{703\,{{k_1}^3}}\over 
    {68992\,{{k_2}^2}\,{{k_3}^4}}} + 
{{19867\,{{k_1}^2}}\over {206976\,k_2\,{{k_3}^4}}}   
	\nonumber  \\
	 &  &  +  
{{5311\,{{k_1}^2}}\over 
    {34496\,{{k_2}^2}\,{{k_3}^3}}} + 
{{42983\,k_1}\over {362208\,k_2\,{{k_3}^3}}}     + 
{{131\,k_1}\over {3696\,{{k_2}^2}\,{{k_3}^2}}} + 
{{28393}\over {19712\,k_1\,k_2\,k_3}}
\nonumber  \\
	 &  &  +   
{{53973\,{{k_1}^7}}\over 
    {1931776\,{{k_2}^5}\,{{k_3}^5}}}+ 
{{108685\,k_1\,k_2}\over {181104\,{{k_3}^5}}}+
{{59599\,{{k_1}^3}}\over 
    {362208\,{{k_2}^3}\,{{k_3}^3}}}
    \nonumber  \\
	 &  &  + {\rm permutations}.
	\label{bispecnm2}
\end{eqnarray}

\noindent A well-known  approximation scheme that provides  
insight into the physics of the non-linear regime is the Zel'dovich 
approximation (ZA) (Zel'dovich 1970). We therefore also consider the 
perturbative expansion for the ZA dynamics (see Appendix~\ref{app:ZA}), 
and calculate one-loop 
corrections to the bispectrum and skewness factor as well (loop 
corrections to the bispectrum 
in the ZA have been recently considered for spectra with small 
scale cutoffs and spectral indices $n>-1$ by Bharadwaj (1996)). For the 
one-loop bispectrum, again for $n=-2$, we find:

\begin{eqnarray}
		 b^{(1)}(k_1,k_2,k_3) & = & -{{27}\over {128\,{{k_1}^3}}} - 
{{9\,k_1}\over {256\,{{k_2}^4}}} + 
{{27}\over {256\,k_1\,{{k_2}^2}}}  +
{{2187\,{{k_1}^7}}\over 
    {32768\,{{k_2}^5}\,{{k_3}^5}}}
	\nonumber  \\
	 &  &  + 
{{9\,{{k_1}^6}}\over {256\,{{k_2}^4}\,{{k_3}^5}}} - 
{{2853\,{{k_1}^5}}\over 
    {16384\,{{k_2}^3}\,{{k_3}^5}}} - 
{{9\,{{k_1}^4}}\over {256\,{{k_2}^2}\,{{k_3}^5}}} - 
{{1467\,{{k_1}^3}}\over {32768\,k_2\,{{k_3}^5}}}
	\nonumber \\
	 &  &  + 
{{2493\,k_1\,k_2}\over {8192\,{{k_3}^5}}}  - 
{{3\,{{k_1}^4}}\over {256\,{{k_2}^3}\,{{k_3}^4}}} + 
{{3\,{{k_1}^2}}\over {256\,k_2\,{{k_3}^4}}}  + 
{{4863\,{{k_1}^3}}\over 
    {16384\,{{k_2}^3}\,{{k_3}^3}}}
	\nonumber  \\
	 &  &  + 
{{3\,{{k_1}^2}}\over {128\,{{k_2}^2}\,{{k_3}^3}}} - 
{{237\,k_1}\over {8192\,k_2\,{{k_3}^3}}}  + 
  {{11823}\over {16384\,k_1\,k_2\,k_3}}    
+ {\rm permutations}. \nonumber  \\
	 &  &
	\label{bispecnm2ZA}
\end{eqnarray}

\noindent Using the one-loop power spectrum for $n=-2$ given 
in SF2  (see also Makino et al. (1992))

\begin{equation}
	p^{(1)}(k) = \frac{55}{98 \ k}
	\label{1loopp},
\end{equation}

\noindent where $P^{(1)}(k) \equiv A^{2} a^{4}
\pi^{3} \ p^{(1)}(k)$, we can obtain the one-loop hierarchical 
amplitude $Q^{(1)}$ 
from Eq.~(\ref{q1l}). Since $Q^{(1)}$ depends on time, a convenient 
parametrization of the degree of nonlinear evolution which takes 
advantage of self-similarity is to write wave-vectors in terms 
of the correlation length $R_{0}$ defined in Eq.~(\ref{R0}), which 
for scale-free initial power spectra and Gaussian smoothing gives

\begin{equation}
	R_{0}^{n+3} \equiv 2\pi \ A \ a^{2} \ \Gamma \Big( \frac{n+3}{2} \Big)
	\label{R0pl}.
\end{equation}

\begin{figure}[t!]
\centering
\centerline{\epsfxsize=11. truecm \epsfysize=10. truecm 
\epsfbox{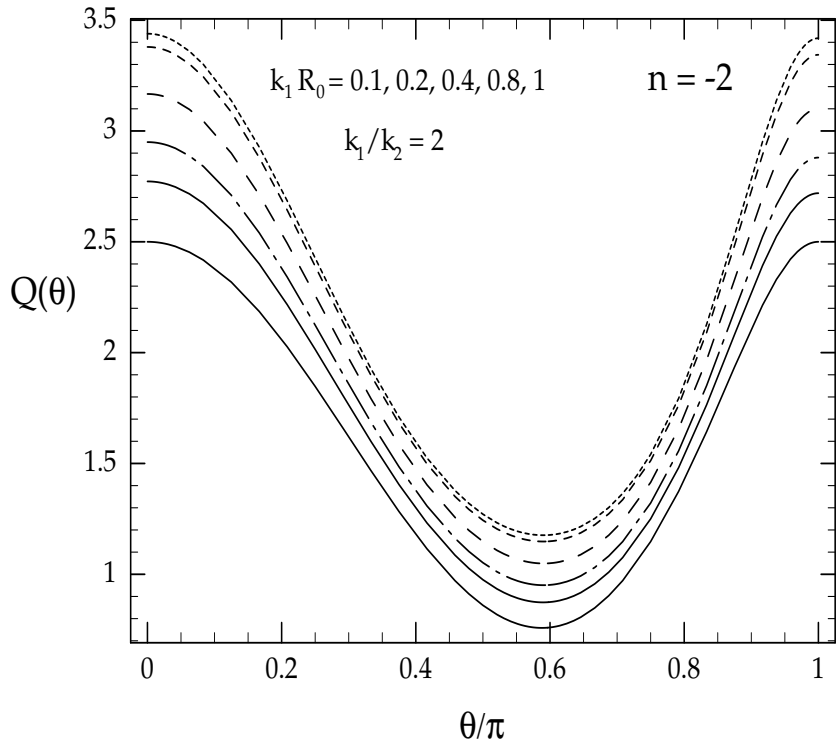}}

\caption{The hierarchical amplitude $Q$ (see Eq.~(\protect{\ref{Qexpand}})) 
for triangle configurations 
($k_{1}=1$, $k_{2}=0.5$; $\hat{k}_{1}\cdot \hat{k}_{2} \equiv 
\cos{\theta}$)
as  a function of the angle $\theta$ to one-loop. The lowest full curve shows 
$Q$ at tree-level, whereas the subsequent curves correspond to 
different stages of non-linear evolution parameterized by the first 
side of the triangle in terms of the correlation length, 
$k_{1} R_{0} =0.1,0.2,0.4,0.8,1$ (from bottom to 
top).
}
\label{figbinm2}
\end{figure}

\begin{figure}[t!]
\centering
\centerline{\epsfxsize=11. truecm \epsfysize=10. truecm 
\epsfbox{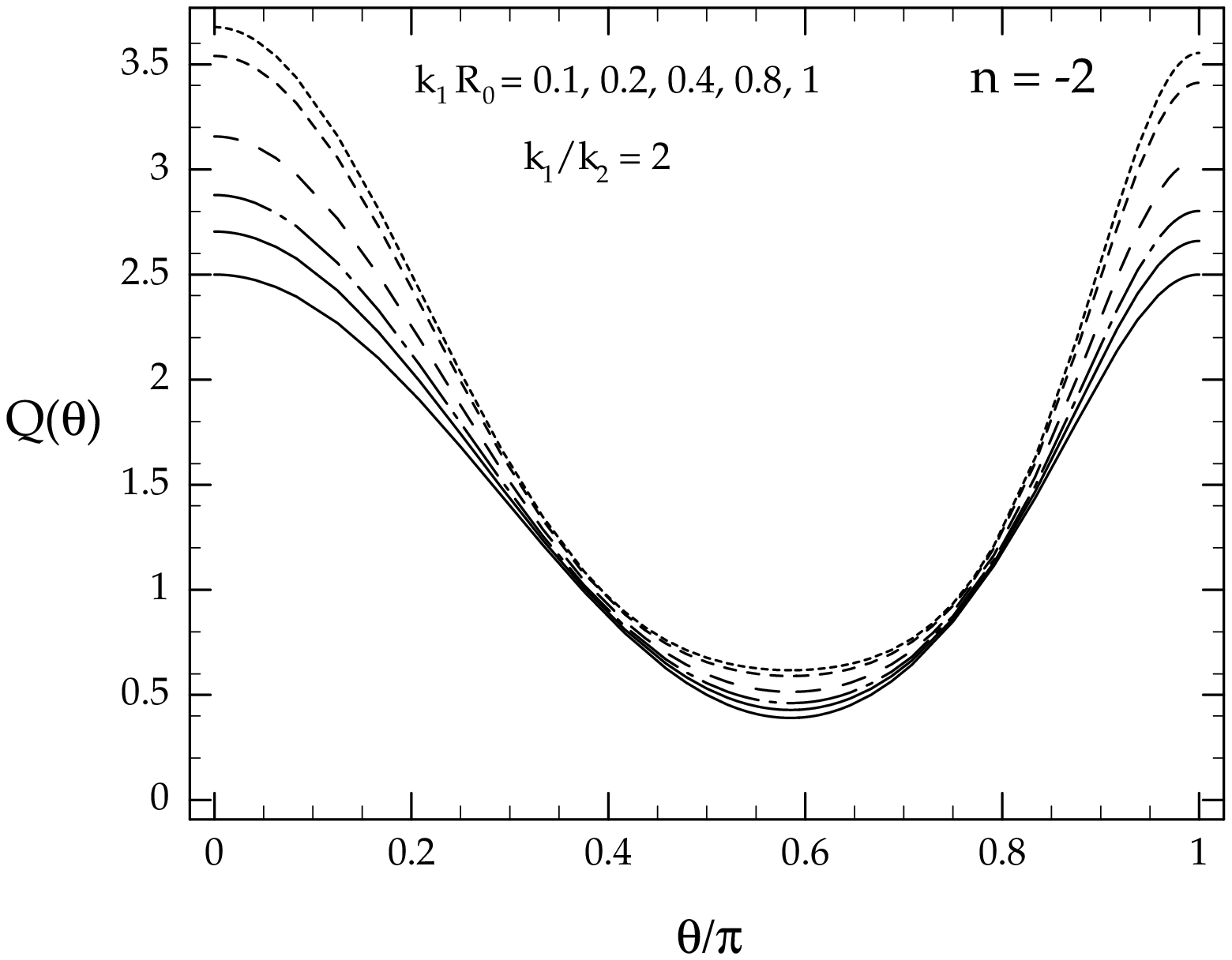}}

\caption{Same as Figure~\protect{\ref{figbinm2}} for the Zel'dovich 
approximation.}
\label{figbizanm2}
\end{figure}

\noindent Figures~\ref{figbinm2} and~\ref{figbizanm2} show
 the resulting hierarchical amplitude $Q$ (see Eq.~(\ref{Qexpand})), 
 for the exact dynamics (ED) and the Zel'dovich approximation (ZA)  
respectively. We see that for the ED one-loop corrections to $Q$ 
are in general not negligible even for weakly nonlinear scales. 
When $k_{1} R_{0} \approx 1$, the contribution to the variance 
per logarithmic interval $\Delta(k) \equiv 4\pi k^{3} P(k)$ becomes of 
order one, and  we 
expect one-loop perturbation theory to break down, since the scales 
considered become comparable to the correlation length. It is  
interesting  that at these scales Eq.~(\ref{Qexpand}) {\it saturates}, 
that is, the one-loop quantities $B^{(1)}$ and $\Sigma^{(1)}$ dominate 
over the corresponding tree-level values and further time evolution 
does not change the amplitude $Q$, because  $B^{(1)}$ and $\Sigma^{(1)}$ 
have the same scale and,  by self-similarity,  time-dependence. 
Note that for this initial spectrum, 
the one-loop correction {\it enhances} the 
configuration dependence of the tree-level bispectrum as the system 
evolves to the non-linear regime. This enhancement is  stronger 
for the ZA, which is understandable in view of the tendency of this 
dynamics to produce highly anisotropic two-dimensional structures 
(pancakes).  Note that the 
ZA underestimates the one-loop correction, in 
correspondence with the unsmoothed skewness (SF1) and power 
spectrum results (SF2).

\begin{figure}[t!]
\centering
\centerline{\epsfxsize=12. truecm \epsfysize=10. truecm 
\epsfbox{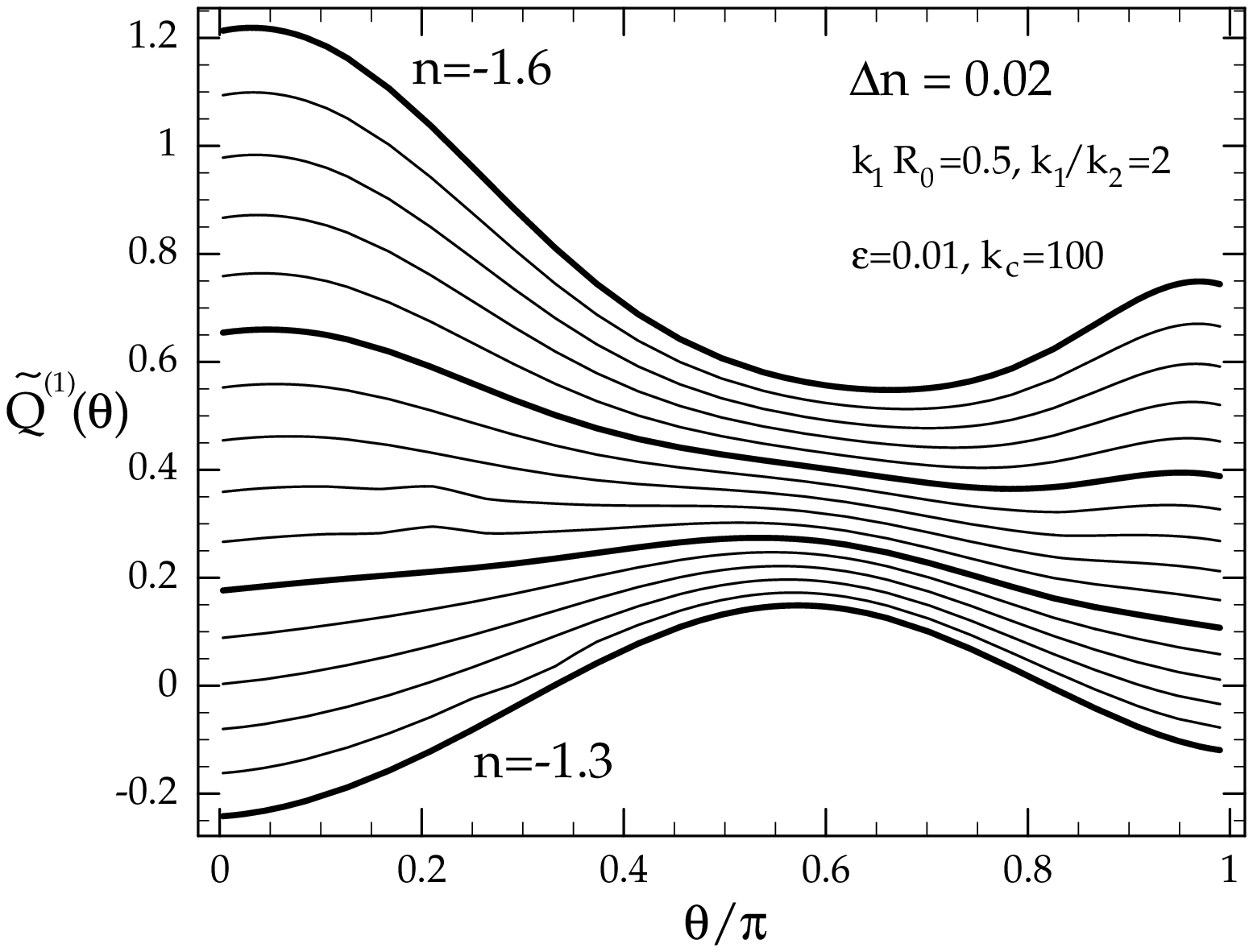}}

\caption{The one-loop hierarchical amplitude $\tilde{Q}^{(1)}$ 
(see Eq.~(\protect{\ref{q1l}}))
for triangle configurations 
($k_{1}=1$, $k_{2}=0.5$; $\hat{k}_{1}\cdot \hat{k}_{2} \equiv 
\cos{\theta}$)
as  a function of the angle $\theta$ for different spectral indices 
$n$. The spectral index runs from $n=-1.6$ (top full curve) to 
$n=-1.3$ (bottom full curve) in steps of $\Delta n =0.02$. This shows 
the transition from positive to negative one-loop corrections as $n$ 
is increased. The linear power spectrum in this figure
is  such that 
${\cal P}_{11}(k) \equiv (k R_{0})^{n} 
\exp[-(\epsilon /k)^{4}] \exp[-(k /k_{c})^{4}]
/[2 \pi \Gamma[(n+3)/2] ]$. 
}
\label{figbifn}
\end{figure}

\begin{figure}[t!]
\centering
\centerline{\epsfxsize=12. truecm \epsfysize=10. truecm 
\epsfbox{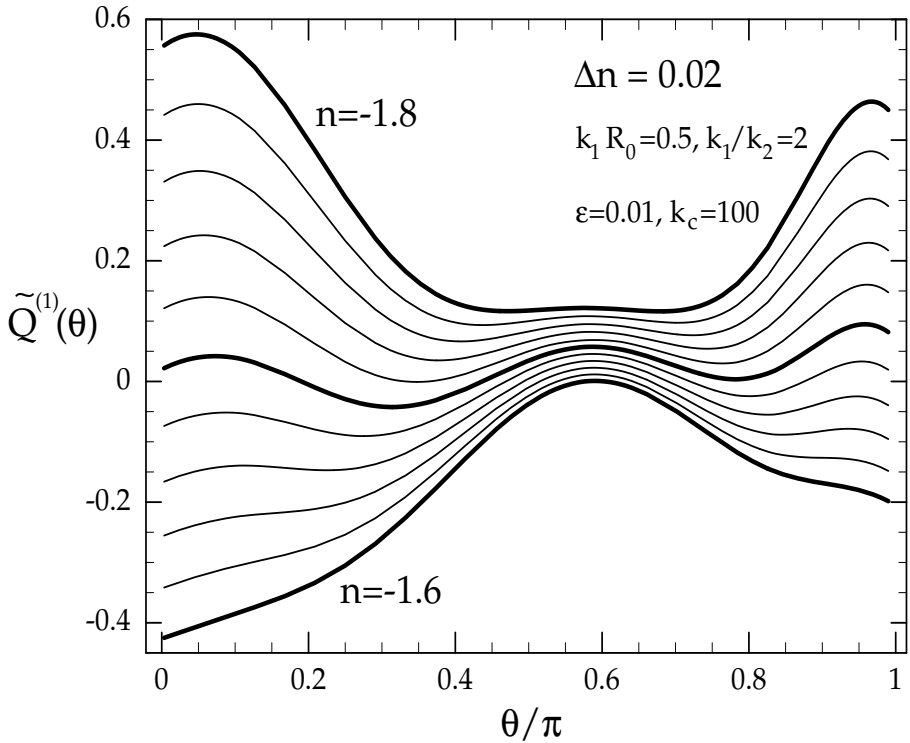}}

\caption{Same as Figure~\protect{\ref{figbifn}} for the Zel'dovich 
approximation.}
\label{figbizafn}
\end{figure}

\begin{figure}[t!]
\centering
\centerline{\epsfxsize=11. truecm \epsfysize=10. truecm 
\epsfbox{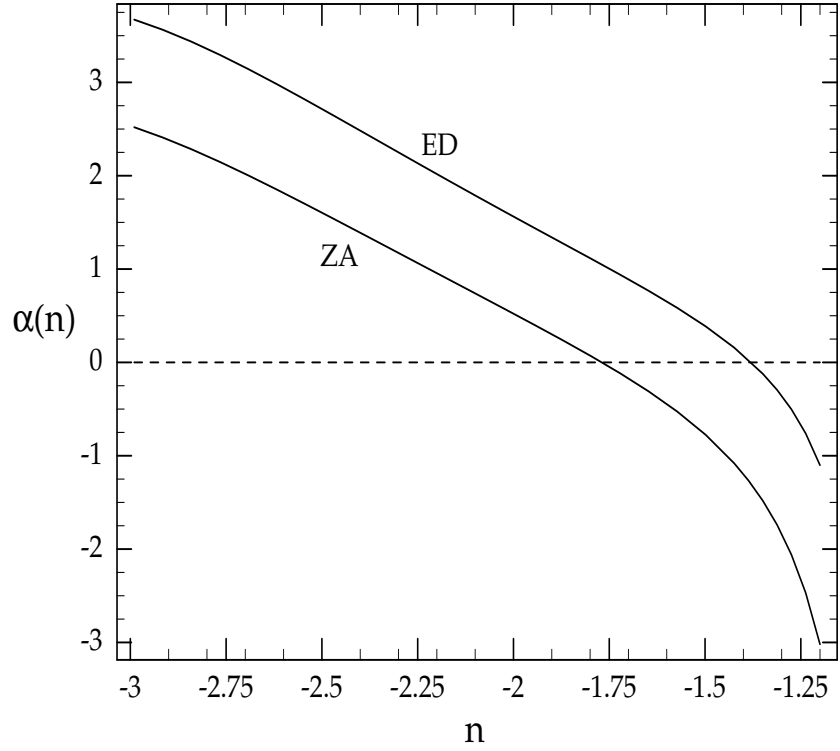}}

\caption{One-loop corrections  to the power spectrum in the exact 
dynamics (ED) and Zel'dovich approximation (ZA) as a function of 
spectral index (see Eq.~(\protect{\ref{Palpha}})).}
\label{alphafn}
\end{figure}

Based on results from N-body simulations, 
it has been pointed out by Fry et al. (1993) (see also Fry et al. (1995))
that for $n=-1$ nonlinear 
evolution tends to ``wash out'' the configuration dependence of the 
bispectrum present at the largest scales (and given by tree-level 
perturbation theory), giving rise to the so-called hierarchical form $Q \approx 
const$ in the strongly non-linear regime. 
One-loop perturbation theory must predict this 
feature in order to be  a good description of the transition to the 
nonlinear regime. To study this, we integrated numerically the one 
loop bispectrum for different spectral indices to understand the transition 
from the behavior at  $n=-2$  to the $n=-1$ spectrum 
(for $n \neq -2$ the one-loop bispectrum 
can be represented 
in terms of hypergeometric functions of two variables 
(see Appendix~\ref{app:dr})). Figure~\ref{figbifn} 
shows the result of such a calculation for the exact dynamics, 
in terms of the one-loop 
hierarchical amplitude $\tilde{Q}^{(1)}$ (see Eq.~(\ref{q1l})) 
for spectral indices running from 
$-1.6$ to $-1.3$. Clearly  one-loop perturbation theory 
predicts a change in behavior of the nonlinear evolution: for $n \la 
-1.4$ the one-loop corrections {\it enhance} the configuration dependence of 
the bispectrum, whereas for $n \ga -1.4$, they tend to cancel it, in 
qualitative agreement with numerical simulations. Note that this 
``critical index'' $n_{c} \approx -1.4$ is the same spectral index at 
which one-loop corrections to the power spectrum vanish, marking the 
transition between faster and slower than linear growth of the 
variance of density field fluctuations (SF2) (see 
also Makino et al (1992), \L okas et al. (1995b), Bagla \& 
Padmanabhan (1996)). Figure~\ref{figbizafn} shows that 
the same situation arises in the Zel'dovich approximation, which has 
$n_{c} \approx -1.75$. Figure~\ref{alphafn} displays the one-loop 
correction to the power spectrum in both the exact dynamics (ED) and 
the Zel'dovich approximation (ZA) in terms of the function $\alpha 
(n)$ defined by

\begin{equation}
	{\cal P}(k R_{0}) \equiv \frac{(k R_{0})^{n}}{2\pi\Gamma[(n+3)/2]}
	\Big[ 1 + \alpha(n) \ (k R_{0})^{n+3}  \Big]
	\label{Palpha},
\end{equation}

\noindent obtained by dimensional regularization in SF2.
Figures~\ref{figbifn}, \ref{figbizafn}, and \ref{alphafn}
clearly illustrate the special character of the 
critical index in both dynamics.  Note that in Figures~\ref{figbifn} 
and \ref{figbizafn}, the calculation is done by numerical integration 
and the linear power spectrum is not exactly scale-free, which can 
account for the very small shift in the critical index in these figures 
with respect to the exact scale-free case in Fig.~\ref{alphafn}.

The change in behavior of the one-loop corrections 
at $n \approx n_c$ can be understood in
physical terms as 
follows. As we increase $n$, the increase in small-scale power
generates 
random motions which tend to prevent the collapse of high density regions 
(the ``previrialization''
effect (Davis \& Peebles 1977, Evrard \& Crone 1992, \L okas et
al. 1995b, Peebles 1990). This is reflected
in the sign change of one-loop corrections to the variance,
which measures the growth of fluctuations. Another manifestation
of this effect, which influences the shape of the 
bispectrum, is that random motions due to small-scale power  
cause structures to be less anisotropic and flows to have a 
smaller
coherence length. This leads, upon ensemble averaging, to a
cancellation of the configuration dependence in the
hierarchical amplitude $Q$ in Fourier space. 
A quantitative 
comparison of the predictions of one-loop perturbation theory with 
N-body simulations for the hierarchical amplitude 
$Q$ is under way and is  the subject of a forthcoming 
paper (Scoccimarro et al. 1996).

\clearpage 

\subsection{Skewness: Comparison with Numerical Simulations}
\label{sec:s3}

We now consider one-loop corrections to the skewness factor. Since 
tree-level one-point cumulants such as the skewness and kurtosis 
are given by the spherical collapse 
dynamics (Bernardeau 1992, Bernardeau 1994b), one-loop contributions 
contain the first 
corrections to the spherical model coming from tidal motions. 
Given the analytic results for the $n=-2$ bispectrum in the previous section,
we can use Eq.(\ref{S3loop}) to integrate numerically the one-loop 
skewness for different window functions. For the exact
dynamics, we obtain (see Eq.~(\ref{S3expand})):

\begin{mathletters}
\label{S3nm2TH}
\begin{eqnarray}
	S_3^{ED}(R) &= & \frac{3.86 + 9.97 \  \sigma^2_{TH}(R)}
	{1 + 1.76 \  \sigma^2_{TH}(R)} \label{S3nm2THa},\\
	& \approx & 3.86 + 3.18 \  \sigma^2_{TH}(R)
	\label{S3nm2THb},
\end{eqnarray}
\end{mathletters}

\noindent for top-hat smoothing, and

\label{S3nm2G}
\begin{equation}
	S_3^{ED}(R) =  \frac{4.02 + 10.91 \  \sigma^2_{G}(R)}
	{1 + 1.76 \  \sigma^2_{G}(R)} 
	 \approx  4.02 + 3.83 \  \sigma^2_{G}(R)
	\label{S3nm2Gb},
\end{equation}

\noindent for Gaussian smoothing. For the Zel'dovich
approximation we get:

\label{S3nm2THza}
\begin{equation}
	S_3^{ZA}(R) =  \frac{3 + 2.59 \  \sigma^2_{TH}(R)}
	{1 + 0.59 \  \sigma^2_{TH}(R)} 
	 \approx  3 + 0.82 \  \sigma^2_{TH}(R)
	\label{S3nm2THzab},
\end{equation}

\noindent for top-hat smoothing, and

\begin{figure}[t!]
\centering
\centerline{\epsfxsize=11. truecm \epsfysize=10. truecm 
\epsfbox{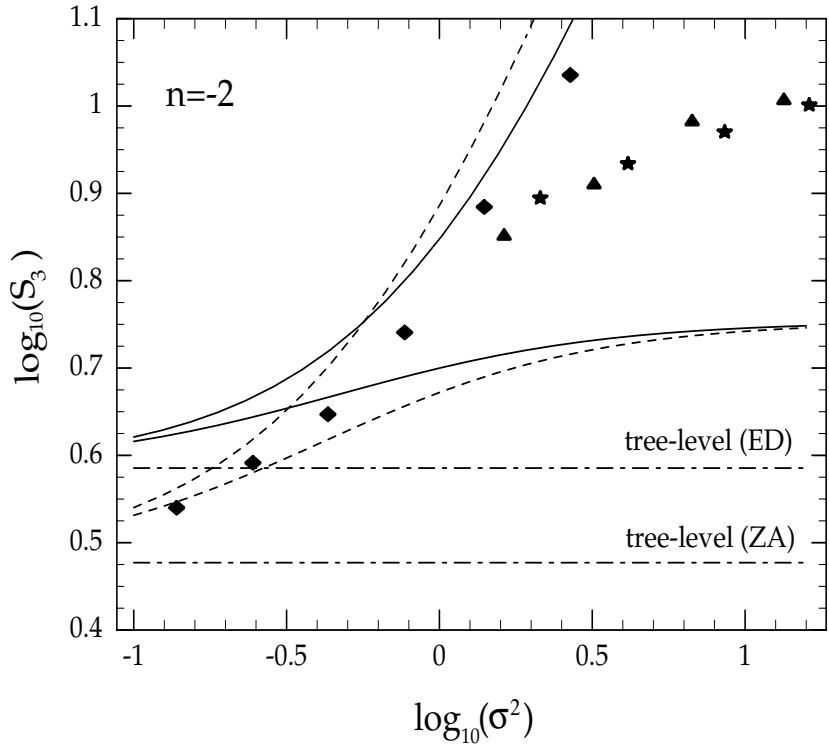}}

\caption{The skewness factor $S_{3}$ 
as  a function of the variance of density fluctuations 
$\sigma^{2}(R)$ for  spectral index $n=-2$. 
The symbols show the results from numerical simulations 
by Colombi et al. (1996) for top-hat smoothing. 
Different symbols correspond
to different output times: diamonds ($a=2$), triangles ($a=3.2$),
and stars ($a=5.2$). Estimated error bars in 
these measurements are $\pm 0.1$ (systematic) in 
logarithmic scale (Colombi et al. 1996). 
The solid curves 
correspond to the prediction for 
top-hat smoothing (TH) of exact one-loop perturbation theory (ED), 
Eqs.~(\protect{\ref{S3nm2THa}}) (bottom) and  
(\protect{\ref{S3nm2THb}})  (top). Dashed lines denote the 
tree-level Zel'dovich approximation (ZA) plus one-loop ED, 
Eqs.~(\protect{\ref{S3nm2THzaeda}}) (bottom) and  
(\protect{\ref{S3nm2THzaedb}})  (top). 
 Dot-dashed lines correspond to tree-level values 
in ED and ZA. }
\label{figS3}
\end{figure}

\label{S3nm2Gza}
\begin{equation}
	S_3^{ZA}(R) =  \frac{3.14+ 2.86 \  \sigma^2_{G}(R)}
	{1 + 0.59 \  \sigma^2_{G}(R)} 
	 \approx  3.14 + 1.00 \  \sigma^2_{G}(R)
	\label{S3nm2Gzab},
\end{equation}

\noindent for Gaussian smoothing. 
We also computed one-loop
corrections to the skewness in the Local Lagrangian
approximation scheme discussed by Protogeros \& Scherrer
(1996). 
We obtain (see
Appendix~\ref{app:lla}):

\label{S3nm2THps}
\begin{equation}
	S_3^{LLA}(R) =  \frac{4 + 9.43 \  \sigma^2_{TH}(R)}
	{1 + 1.39 \  \sigma^2_{TH}(R)} 
	 \approx  4 + 3.87 \  \sigma^2_{TH}(R)
	\label{S3nm2THpsb},
\end{equation}

\noindent for top-hat smoothing. Note that the results in 
Eqs.~(\ref{S3nm2TH})-(\ref{S3nm2THps}) are all for $n=-2$. 
A visual summary of the exact perturbative
results is given in Fig.~\ref{figS3}, where we compare to the
numerical simulations by Colombi et al (1996). These N-body
simulations used a tree code (Hernquist, Bouchet \& Suto 1991), with 
$64^3$ particles in a cubic box with periodic boundary
conditions. Symbols in Fig.~\ref{figS3} correspond
to different output times: diamonds ($a=2$), triangles ($a=3.2$),
and stars ($a=5.2$). The estimated systematic 
uncertainties in these measurements  
of skewness is
$\pm 0.1$ in logarithmic scale, which is due to the uncertainty in the 
finite volume correction applied to the simulation data 
(see Colombi et al. (1996) and also Colombi et al. (1994),
Hivon et al. (1995) for details). This 
correction, which is 
rather large because of the
large-scale power present in the $n=-2$ spectrum, is more important 
when the correlation length becomes a non-negligible fraction of the 
box size (i.e., for larger $a$). For a given time output, 
the finite volume correction is more important for large scales. 
Note that only the
 two latest outputs in this figure have been 
corrected for finite volume effects; for the first output ($a=2$) 
this correction should be negligible.

The two solid curves in Fig.~\ref{figS3} correspond to the 
predictions given in Eq.~(\ref{S3nm2THa}) (bottom) and  
Eq.~(\ref{S3nm2THb})  (top).  The lower curve shows a saturation at 
values of $\sigma^{2} \approx 1 $ such that one-loop corrections in 
Eq.~(\ref{S3nm2THa}) dominate over the tree-level contributions, 
similar to what happens with the hierarchical amplitude $Q$. This 
saturation value, however, is not in good agreement with the numerical 
simulation data, which is not surprising given that those scales are 
well into the non-linear regime. 
We note that the N-body
 results are systematically lower (although within the error
 bars) than the one-loop perturbative
 calculation in the weakly non-linear regime. 
 In fact, as $\sigma^2 \rightarrow 0$ they approach
 asymptotically the tree-level value given in the Zel'dovich
 approximation. This is most likely an artifact coming from the 
 fact that the simulation uses ZA initial
 conditions, which have not been erased by the
 relatively early output time at $a=2$ (Baugh et al. 1995). Note that the 
 dashed curves given by:

\begin{mathletters}
\label{S3nm2THzaed}
\begin{eqnarray}
	S_3(R) &= & \frac{3 + 9.97 \  \sigma^2_{TH}(R)}
	{1 + 1.76 \  \sigma^2_{TH}(R)} \label{S3nm2THzaeda},\\
	& \approx & 3 + 4.69 \  \sigma^2_{TH}(R)
	\label{S3nm2THzaedb},
\end{eqnarray}
\end{mathletters}

\noindent which are the tree-level value given by the ZA
 plus the one-loop correction in the exact dynamics,
fit the numerical results better,  suggesting that indeed transients from 
the ZA initial conditions are still present in the first output.
It is interesting to note that the expansion for large scales given in 
Eq.~(\ref{S3nm2THzaedb}) seems to describe the transition to the 
non-linear regime better than Eq.~(\ref{S3nm2THzaeda}), which soon 
becomes dominated by one-loop corrections and driven to the saturation value.
Overall, however, we see that one-loop
 perturbation theory agrees with the simulation within the error bars 
 even on scales where $\sigma^2 \approx 1$, 
 and therefore describes most of the transition from 
 the tree-level result (valid in the limit $\sigma^{2} \rightarrow 0$) 
 to the nonlinear regime where $S_{3}$ approximately approaches a 
 constant value,  in agreement with the corresponding results for the
 power spectrum (SF2). A more detailed comparison, with
 more accurate N-body measurements, will be presented elsewhere.

The Zel'dovich approximation  for $S_3$ clearly
underestimates the numerical simulation  and the exact dynamics 
perturbative results,  in
agreement with previous results for unsmoothed
fields in SF1. Note that it also fails to describe properly 
the transition to the non-linear regime. 
The Local Lagrangian approximation (LLA) 
does reasonably well, although it overestimates
 the  exact perturbative results; this is not surprising, given that
it has been ``designed'' to reproduce the tree-level one-point
cumulants (Protogeros \& Scherrer 1996). 
On the other hand, a phenomenological model 
recently proposed by Munshi \& Padmanabhan (1996),  
which assumes the hierarchical ansatz (equivalent to a constant
$S_{3}$), 
predicts $S_{3} = 11.76$ for $-0.6 \la \log_{10} \sigma^{2} \la 0.3$, 
in  disagreement with the numerical simulation results, which do 
not  support the hierarchical assumption in the transition to the 
non-linear regime.

It is interesting to note the importance of smoothing in the value of 
one-loop corrections by comparing the above results  to the unsmoothed 
values found in SF1. For the exact dynamics, we obtained for 
unsmoothed fields

\begin{equation}
	S_3^{ED} \approx  4.86 + 10.03 \  \sigma^2
	\label{S3nm2u},
\end{equation}

\noindent for  the $n=-2$ spectrum,  where the corresponding 
result for the Zel'dovich approximation reads:

\begin{equation}
	S_3^{ZA} \approx  4 + 4.69 \  \sigma^2
	\label{S3nm2uZA}.
\end{equation}

\noindent In each case, smoothing reduces the relative importance of 
the loop corrections. For  the exact dynamics, one-loop corrections 
to the smoothed skewness begin to 
dominate over the tree-level contribution for $\sigma^{2} \approx 1$, 
instead at $\sigma^{2} \approx 1/2$ for unsmoothed fields.

Another interesting issue is the spectral  dependence of 
the one-loop corrections to $S_3$. 
One-loop corrections to the variance and average two-point correlation 
function show a linear dependence on spectral index for $-3 < n 
< -1.5$ (SF2), and we conjecture that a similar behavior 
extends to $S_{3}$. 
For $n=-3$, smoothed and 
unsmoothed quantities coincide, because small-scale filtering 
does not affect statistical properties for a model with such 
extreme large-scale power (e.g., the variance is 
infrared-divergent). In this case, we have (SF1) 

\begin{equation}
	S_3(R) \approx  \frac{34}{7} + 10 \  \sigma^2_{TH}(R) 
	\ \ \ \ \ (n=-3)
	\label{S3nm3}.
\end{equation}

\noindent Taking into account the above results for $n=-2$, 
Eq.~(\ref{S3nm2TH}) and assuming linear behavior with $n$, we expect 
that for top-hat smoothing, 

\begin{eqnarray}
	S_3(R,n) & \approx & \frac{34}{7} - (n+3) +  \sigma^2_{TH}(R) \ \Big[
	10 - 6.8 \ (n+3) \Big] \nonumber \\
	& \approx & \Big[ \frac{34}{7} + 10 \ \sigma^2_{TH}(R) \Big] - (n+3) 
	\ \Big[ 1 + 6.8 \ \sigma^2_{TH}(R) \Big] 
	\label{S3n}.
\end{eqnarray}

\noindent  In fact, 
as $n \rightarrow -1.6$, this ansatz  leads to $S_3^{(1)} \rightarrow 0$, 
which is probably outside the region of validity of the linear
extrapolation. For $n=-1$, for example, numerical simulations show that 
$S_3^{(1)} > 0$ (Colombi et al. 1996).

\section{Conclusions}
\label{cha:conc}

We have 
calculated one-loop corrections (non-linear corrections {\it beyond} 
leading order) to the bispectrum and skewness 
of the cosmological density field 
including smoothing effects, induced by gravitational evolution
for Gaussian initial conditions and  scale-free initial power spectra.
These results extend previous calculations done at tree-level and 
allow us  to probe the transition to the non-linear regime.

We have shown that the one-loop bispectrum follows a similar behavior
as a function of 
spectral index as the one-loop power spectrum. For $n \la -1.4$, one-loop 
corrections {\it increase} the configuration dependence of the bispectrum; 
 for $n \ga -1.4$, one-loop corrections tend to cancel the 
configuration dependence of the tree-level bispectrum, in  
agreement with $n=-1$ numerical simulations. Therefore, there is  
a ``critical index'' $n_{c} \approx -1.4$, similar to what happens in 
the case of the power spectrum, where one-loop corrections become 
negative at $n \ga -1.4$ (SF2), indicating a slowing
down of the growth of fluctuations (\L okas et al. 1995b). This increase
in the configuration dependence of the bispectrum 
for $n < n_c $ is a prediction of one-loop
perturbation theory that can be tested against numerical
simulations. The configuration
dependence of the bispectrum is due to the
anisotropy of structures and flows in real space, and therefore has a
direct physical meaning. In this respect, the one-loop 
bispectrum  for
the Zel'dovich approximation, which is well
known to produce highly anisotropic structures (pancakes), shows
a stronger configuration dependence, as expected. 
We interpret the change in
behavior of the bispectrum as the spectral index increases 
as a result of the increased   
effect of small-scale power in the collapse of high density
regions 
(the ``previrialization'' effect (Davis \& Peebles 1977, 
Evrard \& Crone 1992, \L okas et al. 1995b, Peebles 1990). 
The random motions due to small-scale power
tend to slow down the collapse and disrupt coherent
structures and flows on small scales, which is reflected  in the one-loop
corrections to the power spectrum and bispectrum respectively.

For spectral indices $n 
< -1$, self-similarity is retained at the one-loop level, and one can 
calculate loop corrections in the scaling regime by using the 
technique of dimensional regularization. We obtained explicit analytic results
for the one-loop bispectrum for $n=-2$ initial conditions; this allowed 
us to calculate the skewness factor for top-hat and Gaussian 
smoothing. We then extended the results in SF1
to include local averaging of the fields; for $n=-2$, the smoothed one-loop
correction is reduced by more than a factor of 2 from its
unsmoothed value, which shows the importance of smoothing in determining 
the value of one-loop corrections. 
The results for top-hat smoothing compare  well with 
the corresponding measurements in 
numerical simulations, providing a description of most of the transition from 
the tree-level value at large scales ($\sigma \rightarrow 0$) to the  
non-linear regime where $S_{3}$ attains an approximate
 constant value in 
accord with expectations based on stable clustering.

The results presented in this work suggest future directions in
which one can improve the current understanding of the
transition to the non-linear regime. 
One obvious extension
would be to consider spectral indices $n \geq -1$, to generalize
the present results for arbitrary scale-free spectra. This would 
involve taking into account the effects of small-scale
fluctuations on the evolution of large-scale modes in a way
that absorbs the divergences that appear  in the present
formalism (renormalization), and therefore recovers self-similar 
evolution for  statistical quantities such as the power spectrum 
and bispectrum. This would lead to a better
understanding of the role of previrialization in  determining
the structure of the correlation functions on intermediate scales. 
We hope to come back to this point in the near future.

Since realistic power spectra are not scale-free, an 
important further step  is to consider initial conditions such as
those given by the cold dark matter (CDM) model and its variants. Since these
models have effective spectral indices in the range 
$n_{eff} \approx -2$~to~$-1$
over the scales of interest, we expect that
they will show similar features to the ones we
presented here.  Nevertheless  
explicit calculations are required in
order to assess the effect of one-loop corrections in the
determination of bias from the configuration dependence of the 
bispectrum (Fry 1994). Similarly, recent claims 
by Jing \& B\"{o}rner (1996) that there is a
discrepancy in the three-point function between 
tree-level perturbation theory and  numerical
simulations for CDM models, may be properly addressed by taking
into account one-loop corrections. As we showed in this
work, these can be non-negligible even on weakly non-linear scales,
depending on the initial spectrum. Work is in progress on these issues 
(Scoccimarro et al. 1996).

There is clearly much more work to do to understand 
non-linear clustering in an expanding
Universe. However, the interplay between perturbation theory
and N-body simulations suggests that
there are three distinct regimes that  describe its most
important statistical features. At the largest scales, tree-level
perturbation theory is well established as providing a good
description of the correlation functions and the $\sigma \rightarrow 0$
limit of the $S_p$ parameters (Fry 1984, Bernardeau 1994).  
In the strongly non-linear regime, numerical
simulations (Hamilton et al 1991,  
Peacock \& Dodds 1994, Suto 1993, Jain et al. 1995, 
Colombi et al. 1996, Jain 1996) have 
shown reasonable 
agreement with the stable clustering hypothesis, although there
are still large uncertainties due to 
limitations in dynamic range.  In this regime,  there is as yet 
no compelling analytic model  which makes predictions in
good agreement with the numerical results, except probably for 
the two-point function (Sheth \& Jain 1996). In particular, there is no
understanding of the hierarchical structure of the $S_p$
parameters, which seem to reach a plateau in the highly
non-linear regime, $S_p \approx $ constant. 
Finally, the results presented in this work suggest that the third
regime, the transition to the non-linear regime, with $\sigma \approx 
1$, can be understood by one-loop perturbation theory for models 
without excessive small-scale power. This is clearly
promising and deserves further investigation.

\acknowledgments

It's a pleasure to thank  Josh Frieman, who has 
guided me in the research described in this work.
I am indebted to Stephane Colombi for 
providing me with the results of his measurements in numerical 
simulations used in this paper and for many  discussions on this 
subject. Special thanks are  due to Jim Fry, with whom I cross-checked 
my numerical results on the one-loop bispectrum. 
I am also grateful to  R.~Cebral, D.~Chung, L.~Kadanoff, S.~Meyer, 
and M.~Turner for 
comments and discussions.  Financial support by 
the DOE at Chicago and Fermilab and by NASA grant NAG5-2788 at 
Fermilab is particularly acknowledged.

\appendix
\section{Dimensional Regularization} 
\label{app:dr}

To obtain the behavior of the one-loop N-point spectra for $n <-1$, one
can 
use dimensional regularization (see e.g. Collins (1984))
 to simplify considerably the calculations. 
 Since we are interested in the limit 
$k_c \rightarrow \infty$,
all the integrals run from $0$ to $\infty$, and  divergences are 
regulated by changing the dimensionality $d$ of space: 
we set $d = 3 + \varepsilon$ and expand 
in $\varepsilon \ll 1$. For the bispectrum, we need  
the following one-loop three-point integral:

\begin{equation}
J(\nu_1,\nu_2,\nu_3) \equiv 
\int  \frac{d^d {\bf q}}{(q^2)^{\nu_1} [({\bf k_{1}}-{\bf q})^2]^{\nu_2}
[({\bf k_{2}}-{\bf q})^2]^{\nu_3}} 
\label{J}.
\end{equation}

\noindent When one of the indices vanishes, 
e.g. $\nu_{3}=0$, this reduces  to the 
standard formula for dimensional-regularized 
two-point integrals (Smirnov 1991):

\begin{equation}
 J(\nu_1,\nu_2,0) =
 \frac{\Gamma (d/2 -\nu_1)
\Gamma (d/2 -\nu_2)
\Gamma (\nu_1 + \nu_2 - d/2)}{\Gamma (\nu_1) \Gamma (\nu_2) \Gamma (d-
\nu_1 - \nu_2)} \ \pi^{d/2}  \ k_{1}^{d-2 \nu_1 -2 \nu_2}
\label{I}.
\end{equation}

\noindent The integral $J(\nu_1,\nu_2,\nu_3)$ appears in triangle 
 diagrams for massless particles 
 in quantum field theory, and can be evaluated for 
arbitrary values of its parameters in terms of hypergeometric 
functions of two variables (Davydychev 1992). The result is:

 \begin{eqnarray}
 	J(\nu_1,\nu_2,\nu_3) & = & \frac{\pi^{d/2} k_{1}^{d-2 \nu_{123}
 	}}{\Gamma (\nu_1) \Gamma (\nu_2)  \Gamma (\nu_3) \Gamma (d-
    \nu_{123})} \times \ \Bigg( \Gamma (\nu_3) 
    \Gamma (\nu_{123} - d/2) 
 	\nonumber   \\
 	 &  & \times {\rm F}_{4} (\nu_3,\nu_{123} - d/2;1 +\nu_{23}- 
 	 d/2, 1 + \nu_{13}- d/2 ; x , y)
 	\nonumber  \\
 	 &  & \times \Gamma (d/2-\nu_{13}) \Gamma (d/2-\nu_{23}) +
 	 y^{d/2-\nu_{13}} \Gamma (\nu_2) \Gamma (d/2-\nu_1)
 	\nonumber  \\
 	 &  & \times {\rm F}_{4} (\nu_2,d/2 - \nu_1;1 +\nu_{23}- 
 	 d/2, 1 -\nu_{13}+ d/2 ; x , y)
 	\nonumber  \\ 
 	 &  & \times \Gamma (\nu_{13}-d/2) \Gamma (d/2-\nu_{23})
 	 + x^{d/2-\nu_{23}} \Gamma (\nu_1) \Gamma (d/2-\nu_2)
 	\nonumber  \\
 	 &  & \times {\rm F}_{4} (\nu_1,d/2 - \nu_2;1 -\nu_{23}+ 
 	 d/2, 1 + \nu_{13}- d/2 ; x , y)
 	\nonumber  \\ 
 	 &  & \times \Gamma (d/2-\nu_{13}) \Gamma (\nu_{23}-d/2) 
 	 + x^{d/2-\nu_{23}} y^{d/2-\nu_{13}} \Gamma (d/2-\nu_3)
 	\nonumber  \\ 
 	 &  & \times 
 	 {\rm F}_{4} (d-\nu_{123},d/2 - \nu_3;1 -\nu_{23}+ 
 	 d/2, 1 -\nu_{13}+ d/2 ; x , y)
 	\nonumber  \\ 
 	 &  & \times \Gamma (d-\nu_{123}) \Gamma (\nu_{23}-d/2)
 	 \Gamma (\nu_{13}-d/2) \Bigg)
 	\label{Jresult},
 \end{eqnarray}

\noindent where $\nu_{123} \equiv \nu_{1}+\nu_{2}+\nu_{3}$,
$\nu_{ij} \equiv \nu_{i}+ \nu_{j}$, 
$x \equiv ({\bf k}_{2}-{\bf k}_{1})^{2}/k_{1}^{2}$, 
$y \equiv k_{2}^{2}/k_{1}^{2}$, and ${\rm F}_{4}$ is Apell's 
hypergeometric function of two variables, with the series expansion:

\begin{equation}
	{\rm F}_{4} (a,b;c,d;x,y) = \sum_{i=0}^{\infty} \sum_{j=0}^{\infty}
     \frac{x^{i}y^{j}}{i!\ j!} \ \frac{ (a)_{i+j} (b)_{i+j}}{(c)_{i}
     (d)_{j}}
	\label{F4exp},
\end{equation}

\noindent  where $(a)_{i} \equiv \Gamma (a+i) / \Gamma (a)$ denotes the 
Pochhammer symbol.
When the spectral index is $n=-2$, the hypergeometric 
functions reduce to polynomials in their variables due to the 
following useful property for $-a$ a positive integer:

\begin{equation}
	{\rm F}_{4} (a,b;c,d;x,y) = \sum_{i=0}^{-a} \sum_{j=0}^{-a-i}
     \frac{x^{j}y^{i}}{j!\ i!} \ \frac{  (b)_{i+j}}{(c)_{i}
     (d)_{j}} \ \frac{(-1)^{i+j} (-a)!}{(-a-i-j)!}
	\label{F4pol}.
\end{equation}

When using Eq.~(\ref{Jresult}), divergences appear as poles in the 
gamma functions; these can be handled by  the following expansion 
($n=0,1,2, \ldots$ and $\varepsilon \rightarrow 0$):

\begin{equation}
	\Gamma(-n + \varepsilon) = \frac{(-1)^n}{n!} \left[ \frac{1}{\varepsilon} + 
	\psi (n+1) + \frac{\varepsilon}{2} \left( \frac{\pi^2}{3} + \psi^2 (n+1) - 
	\psi' (n+1) \right) + {\cal O} (\varepsilon^2) \right]
	\label{gammapoles},
\end{equation}

\noindent where $\psi (x) \equiv d \ln \Gamma (x) /dx$ and

\begin{eqnarray}
	\psi (n+1) & = & 1 + \frac{1}{2} + \ldots +  \frac{1}{n} - \gamma_e
	\label{psi},  \\
	\psi' (n+1) & = & \frac{\pi^2}{6} - \sum_{k=1}^{n} \frac{1}{k^2}
	\label{psiprime},
\end{eqnarray}

\noindent with $\psi (1) = - \gamma_e = -0.577216 \ldots$
 and $\psi' (1) = \pi^2 /6$.

\section{Zel'dovich Approximation} 
\label{app:ZA}

In this approximation (Zel'dovich 1970, Shandarin \& Zel'dovich
1989), 
the motion of each 
particle is given by its initial Lagrangian displacement. In Eulerian 
space, this is equivalent to replacing the Poisson equation by the 
ansatz (Munshi \& Starobinski 1994):

\begin{equation}
	{\bf v}({\bf x},\tau) = -\frac{2}{3 {\cal H}(\tau) }  \nabla 
	\Phi({\bf x},\tau)
	\label{zel-ansatz},
\end{equation} 


\noindent which is the relation between velocity and gravitational 
potential valid 
in linear theory. The important point about the ZA  
is that a small perturbation in Lagrangian 
fluid element paths carries a large amount of non-linear information
about the  corresponding Eulerian quantities, since the
Lagrangian picture is intrinsically non-linear in the density field. This 
leads to non-zero Eulerian perturbation theory kernels at every order. 
The ZA 
works reasonably well as long as streamlines of flows do not cross each 
other. However,  multistreaming develops at the location of pancakes,  
leading to the breakdown of ZA (Shandarin \& Zel'dovich 1989). The equations 
of motion in Fourier space are:


\label{zeleqsfourier}
\begin{equation}
	{\partial \tilde{\delta}({\bf k},\tau) \over{\partial \tau}} + 
	\tilde{\theta}({\bf k},\tau) = - \int d^3k_1 \int d^3k_2 \delta_D({\bf k} 
	- {\bf k}_1 - {\bf k}_2) \alpha({\bf k}, 
	{\bf k}_1) \tilde{\theta}({\bf k}_1,\tau) 
	\tilde{\delta}({\bf k}_2,\tau)
	\label{zelddtdelta},
\end{equation} 
\begin{equation}
	{\partial \tilde{\theta}({\bf k},\tau) \over{\partial \tau}} - 
	\frac{{\cal H}(\tau)}{2} \ \tilde{\theta}({\bf k},\tau)  = 
	- \int d^3k_1 \int d^3k_2 \delta_D({\bf k} 
	- {\bf k}_1 - {\bf k}_2) \beta({\bf k}, {\bf k}_1, {\bf k}_2) 
	\tilde{\theta}({\bf k}_1,\tau) \tilde{\theta}({\bf k}_2,\tau)
	\label{zelddtthet} ~~.
\end{equation} 

\noindent These equations, together with the perturbative 
expansion~(\ref{ptansatz}), lead to the recursion relations ($n \geq 2$):


\label{zelrecrel}
\begin{eqnarray}
 F_n^{Z}({\bf q}_1,  \ldots  ,{\bf q}_n) & = & \sum_{m=1}^{n-1}
  G_m^{Z}({\bf q}_1,  \ldots  ,{\bf q}_m)  
  \biggl[  \frac{\alpha ({\bf k},{\bf k}_1)}{n}  F_{n-m}^{Z}({\bf q}_{m+1},  \ldots  
 ,{\bf q}_n) +\frac{ \beta({\bf k},{\bf k}_1, {\bf k}_2)}{n(n-1)}  
 \nonumber \\
 & &  \ \ \ \ \ \ \ \ \ \ \ \ \ \ \ \ \ \ \ \ \ \ \ \ \ \ \ \  
\times G_{n-m}^{Z}({\bf q}_{m+1},  \ldots  ,{\bf q}_n) \biggr] 
 \label{ec:FZn},
\end{eqnarray} 

\begin{equation}
	 G_n^{Z}({\bf q}_1,  \ldots  ,{\bf q}_n) = \sum_{m=1}^{n-1}
  G_m^{Z}({\bf q}_1,  \ldots  ,{\bf q}_m)  
 \frac{\beta({\bf k},{\bf k}_1, {\bf k}_2)}{(n-1)}  G_{n-m}^{Z}({\bf q}_{m+1},  
 \ldots   ,{\bf q}_n) 
	\label{ec:GZn},
\end{equation} 

\noindent After symmetrization, 
the density field kernel takes the simple form (Grinstein \&
Wise 1987):

\begin{equation}
 F_n^{Z (s) }({\bf q}_1,  \ldots  ,{\bf q}_n) = \frac{1}{n!} 
 \frac{({\bf k} \cdot {\bf q}_1)}{q_1^2} \ldots 
 \frac{({\bf k} \cdot {\bf q}_n)}{q_n^2}  
 \label{ec:FZ}.
\end{equation}

\noindent  Using Eq.~(\ref{ec:FZ}) for $n=2$ and (\ref{S3tree}), 
one can can calculate the skewness at 
tree-level for different window functions. For top-hat smoothing, the 
result is (Bernardeau 1994)

\begin{equation}
	S_3^{(0)} =  4 - (n+3) 
	\label{S3nTHza},
\end{equation}

\noindent whereas for 
Gaussian smoothing, using the methods described 
by \L okas et al. (1995), we obtain:

\begin{eqnarray}
S_{3}^{(0)} &=& 4  \ {_{2}{\rm F}_{1}} \Big(\frac{n+3}{2},\frac{n+3}{2},
\frac{3}{2},\frac{1}{4}\Big) - (n+3) \  
 {_{2}{\rm F}_{1}} \Big(\frac{n+5}{2},\frac{n+3}{2},
\frac{5}{2},\frac{1}{4}\Big) \nonumber \\
& & + \frac{(n+3)^2}{30} \  
 {_{2}{\rm F}_{1}} \Big(\frac{n+5}{2},\frac{n+5}{2},
\frac{7}{2},\frac{1}{4}\Big) 
	\label{S3treeGza},
\end{eqnarray}

\noindent where ${_{2}{\rm F}_{1}}$ denotes a hypergeometric function. 
More explicitly, for Gaussian smoothing,   
$S_{3}^{(0)}= 3.14, \ 2.80, \ 2.51, \ 2.26,$ $ \ 2.04$ for $n=-2,-1.5, -1, -0.5, 
0$ respectively.

\section{Local Lagrangian Approximation} 
\label{app:lla}

In this approximation (Protogeros \& Scherrer 1996), 
the final density at a Lagrangian 
point ${\bf q}$ at time $\tau$ is assumed to be a function only of the 
initial density at the same Lagrangian point and time $\tau$:

\begin{equation}
	\eta({\bf q},\tau) \equiv \frac{\eta_{0}({\bf q})}{
	[ 1 - a(\tau) \ \delta_{0}({\bf q}) /\alpha ]^{\alpha}} 
	\label{eta},
\end{equation}

\noindent where $\eta \equiv 1 + \delta$, with $\delta_{0}({\bf q}) 
\equiv \delta({\bf q},\tau_{0})$ and $\tau_{0}$ is the initial time. 
The constant $\alpha$ takes the 
value $\alpha=1$ for the planar approximation (which becomes exact 
for one-dimensional collapse), whereas $\alpha=3$ corresponds to  
spherical collapse in the Zel'dovich approximation. Local Lagrangian 
approximations at the level of the equations of motion for fluid 
elements have been considered recently by Hui \& Bertschinger
(1996). 
In this work we 
restrict ourselves to the case $\alpha=3/2$, which, although it has no particular 
physical meaning, can be shown to closely approximate the 
hierarchical amplitudes $S_{p}$ of tree-level perturbation theory for the 
exact dynamics (Bernardeau 1992).  Upon normalization of the 
probability distribution function for $\delta$, 
the local Lagrangian approximation in this 
case reads:

\begin{equation}
	\eta({\bf q},\tau) \equiv \frac{ \left\langle 
	[ 1 - 2 \ \delta_{1}({\bf q},\tau) /3 ]^{3/2}  \right\rangle_{lag}}{
	[ 1 - 2 \ \delta_{1}({\bf q},\tau) /3 ]^{3/2}} 
	\label{etanor},
\end{equation}

\noindent where $< >_{lag}$ denotes ensemble averaging in Lagrangian 
space, and $\delta_{1}({\bf q},\tau)$ corresponds to the 
evolution of the density contrast in linear perturbation theory. The 
$S_{p}$ parameters are defined as ($p >2$)

\begin{equation}
	S_{p} \equiv \frac{ < \delta^{p}  >}{
	<\delta^{2}>^{p-1}} 
	\label{Sp},
\end{equation}

\noindent where the angular brackets correspond to Eulerian ensemble 
averages. For powers of $\eta$ we have (Protogeros \& Scherrer 1996)

\begin{equation}
	< \eta^{m}  > = < \eta^{m-1}  >_{lag}
	\label{eu-lag}.
\end{equation}

\noindent Given that, for Gaussian initial conditions, 
$<\delta_{1}^{m}> = (m-1)!! \ \sigma^{m}$ for $m$ even and 
zero otherwise, one has everything needed to compute Eq.~(\ref{Sp}) 
for unsmoothed density fields. The effects of top-hat
smoothing for power-law power spectra 
can be included by the implicit 
mapping (Bernardeau 1994b, Protogeros \& Scherrer 1996)

\begin{equation}
	\eta_{s} = f[ \delta_{1} \ \eta_{s}^{-(n+3)/6} ]
	\label{etas},
\end{equation}

\noindent where $f(x) \equiv (1 -2x/3)^{-3/2}$ denotes the unsmoothed 
mapping. We are particularly interested in $n=-2$, for which 
Eq.~(\ref{etas}) yields:

\begin{equation}
	\eta_{s} = z^{-6}(\delta_{1}) < z^{6}(\delta_{1}) >_{lag} 
	\label{etasnm2},
\end{equation}

\noindent where $z(\delta_{1})$ denotes the appropriate 
solution to the quartic 
equation $z^{4}=1- 2 \delta_{1}z/3$. Using Eqs.~(\ref{Sp}) and 
(\ref{eu-lag}), we obtain for the skewness and kurtosis factors 

\begin{equation}
	S_{3}(R) = \frac{4 + 509/54 \ \sigma^{2}_{TH}(R)}
	{1 + 25/18 \ \sigma^{2}_{TH}(R)} \approx 
	4 + \frac{209}{54}\ \sigma^{2}_{TH}(R) \approx 
	4 + 3.87 \ \sigma^{2}_{TH}(R)
	\label{S3nm2psTH},
\end{equation}

\begin{eqnarray}
	S_{4}(R) &= &\frac{269/9 + 53661191/373248 \ 
	\sigma^{2}_{TH}(R)}{1 + 25/12 \ \sigma^{2}_{TH}(R)} 
	\approx \frac{269}{9} + \frac{30419591}{373248}\ 
	\sigma^{2}_{TH}(R) \nonumber \\ 
	& \approx & 
	29.88 + 81.49 \ \sigma^{2}_{TH}(R)
	\label{S4nm2THps}.
\end{eqnarray}

\end{document}